\documentclass{aastex62}
\pdfoutput=1
\usepackage{threeparttable} 
\usepackage{verbatim}
\usepackage[T1]{fontenc}
\usepackage[utf8]{inputenc}
\usepackage{makecell}

\graphicspath{{./}{figures/}}

\shorttitle{The 0.8-4.5$\mu$m  broadband transmission spectra of TRAPPIST-1 planets}
\shortauthors{E. Ducrot et al.}

\begin{document}
\title{The 0.8-4.5$\mu$m  broadband transmission spectra of TRAPPIST-1 planets}

\correspondingauthor{Elsa Ducrot}
\email{educrot@uliege.be}

\author{E. Ducrot}
\affil{Space Sciences, Technologies and Astrophysics Research (STAR) \\	Institute, Université de Liège, Allée du 6 Août 19C, B-4000 Liége, Belgium}
\author{M. Sestovic}
\affiliation{University of Bern, Center for Space and Habitability, Gesellschaftsstrasse 6, CH-3012, Bern, Switzerland}
\author{B. M. Morris}
\affiliation{ Astronomy Department, University of Washington, Seattle, WA 98195 USA}

\author{M. Gillon}
\affil{Space Sciences, Technologies and Astrophysics Research (STAR) \\	Institute, Université de Liège, Allée du 6 Août 19C, B-4000 Liége, Belgium}

\author{A. H. M. J. Triaud}
\affiliation{School of Physics \& Astronomy, University of Birmingham, Edgbaston, Birmimgham B15 2TT, UK}
\author{J. de Wit}
\affiliation{Department of Earth, Atmospheric and Planetary Science, \\Massachusetts Institute of Technology, 77 Massachusetts Avenue, Cambridge, MA 02139, USA}
\author{D. Thimmarayappa}
\affil{Space Sciences, Technologies and Astrophysics Research (STAR) \\	Institute, Université de Liège, Allée du 6 Août 19C, B-4000 Liége, Belgium}1
\author{E. Agol}
\affiliation{ Astronomy Department, University of Washington, Seattle, WA 98195 USA}
\author{Y. Almleaky}
\affiliation{Space and Astronomy Department, Faculty of Science, King Abdulaziz University, 21589 Jeddah,
Saudi Arabia}
\affiliation{King Abdullah Centre for Crescent Observations and Astronomy (KACCOA), Makkah Clock,
Saudia Arabia}
\author{A. Burdanov}
\affil{Space Sciences, Technologies and Astrophysics Research (STAR) \\	Institute, Université de Liège, Allée du 6 Août 19C, B-4000 Liége, Belgium}
\author{A. J. Burgasser}
\affiliation{Center for Astrophysics and Space Science, University of California San Diego, La Jolla, CA, 92093, USA}
\author{L. Delrez}
\affiliation{Cavendish Laboratory, JJ Thomson Avenue, Cambridge, CB3 0H3, UK }
\author{B-O. Demory}
\affiliation{University of Bern, Center for Space and Habitability, Gesellschaftsstrasse 6, CH-3012, Bern, Switzerland}
\affil{University of Copenhagen, Centre for Star and Planet Formation, Niels Bohr Institute and Natural History Museum, DK-1350, Copenhagen, Denmark}

\author{E. Jehin}
\affil{Space Sciences, Technologies and Astrophysics Research (STAR) \\	Institute, Université de Liège, Allée du 6 Août 19C, B-4000 Liége, Belgium}
\author{J. Leconte}
\affiliation{Laboratoire d’astrophysique de Bordeaux, Univ. Bordeaux, CNRS, B18N, Allée Geoffroy Saint-Hilaire, F-33615 Pessac, France}
\author{J. McCormac}
\affiliation{Department of Physics, University of Warwick, Gibbet Hill Road, Coventry, CV4 7AL}
\author{C. Murray}
\affiliation{Cavendish Laboratory, JJ Thomson Avenue, Cambridge, CB3 0H3, UK }
\author{D. Queloz}
\affiliation{Cavendish Laboratory, JJ Thomson Avenue, Cambridge, CB3 0H3, UK }
\author{F. Selsis}
\affiliation{Laboratoire d’astrophysique de Bordeaux, Univ. Bordeaux, CNRS, B18N, Allée Geoffroy Saint-Hilaire, F-33615 Pessac, France}
\author{S. Thompson}
\affiliation{Cavendish Laboratory, JJ Thomson Avenue, Cambridge, CB3 0H3, UK }
\author{ V. Van Grootel}
\affil{Space Sciences, Technologies and Astrophysics Research (STAR) \\	Institute, Université de Liège, Allée du 6 Août 19C, B-4000 Liége, Belgium}



\begin{abstract}

{The TRAPPIST-1 planetary system represents an exceptional opportunity for the atmospheric characterization of temperate terrestrial exoplanets with the upcoming James Webb Space Telescope (JWST). Assessing the potential impact of stellar contamination on the planets’ transit transmission spectra is an essential precursor step to this characterization. Planetary transits themselves can be used to scan the stellar photosphere and to constrain its heterogeneity through transit depth variations in time and wavelength. In this context, we present our analysis of 169 transits observed in the optical from space with K2 and from the ground with the SPECULOOS and Liverpool telescopes. Combining our measured transit depths with literature results gathered in the mid/near-IR with Spitzer/IRAC and HST/WFC3, we construct the broadband transmission spectra of the TRAPPIST-1 planets over the 0.8-4.5 $\mu$m spectral range. While planets b, d, and f spectra show some structures at the 200-300ppm level, the four others are globally flat. Even if we cannot discard their instrumental origins, two scenarios seem to be favored by the data: a stellar photosphere dominated by a few high-latitude giant (cold) spots, or, alternatively, by a few small and hot (3500-4000K) faculae. In both cases, the stellar contamination of the transit transmission spectra is expected to be less dramatic than predicted in recent papers. Nevertheless, based on our results, stellar contamination can still be of comparable or greater order than planetary atmospheric signals at certain wavelengths. Understanding and correcting the effects of stellar heterogeneity therefore appears essential to prepare the exploration of TRAPPIST-1’s with JWST.

%
%
}
\end{abstract}
               
%


\section{Introduction}

   The nearby ($\sim$12 pc) TRAPPIST-1 system is composed of an M8-type dwarf star orbited by seven nearly Earth-sized, temperate, planets \citep[][hereafter G17]{Gillon2017}. 
   Considering their transiting nature combined with the infrared brightness ($K$=10.3) and the Jupiter-like size of their host star \citep[$\sim$0.12 $R_\odot$, ][]{VanGrootel2018}, these planets are particularly promising candidates for the first thorough atmospheric characterizations of temperate terrestrial worlds with the upcoming \textit{James Webb Space Telescope} (JWST) \citetext{G17, \citealt{Barstow2016}, \citealt{Morley2017}}. However,  some recent works proposed that an inhomogeneous stellar photosphere -as anticipated for red dwarfs like TRAPPIST-1- could strongly complicate the information content of the exoplanets' transmission spectra, limiting the deciphering of their atmospheric properties \citep[][hereafter R18]{Apai,Rackham2018}. Therefore, the quantification and the correction of this spectral contamination should be a critical preliminary step before any  intensive follow-up of the planets with JWST.
   
From TRAPPIST-1's K2 variability,  R18 estimated TRAPPIST-1's coverage to be $8_{-7}^{+18}\%$ of cold spots and $54_{-46}^{+16}\%$ of hot faculae, assuming Solar-type spots (which maximize the impact on the planets' transit spectra). They concluded that such a strong heterogeneous photosphere could alter the transit depth of the planets by roughly 1 to 15 times the strength of planetary features, dramatically complicating the follow-up observations with JWST. More recently, \citet[][hereafter Z18]{Zhang2018} analyzed the near-IR data obtained with HST/WFC3 for several TRAPPIST-1 planets, and compared their resulting transit spectra with the R18 stellar contamination model. They concluded that the star  should be almost entirely covered by spots ($\sim 30\%$) and faculae ($\sim 63\%$) -essentially  a "two-component photosphere"- and predicted dramatic (a few dozens of \%) chromatic variations of the transit depths, especially in the optical.
 
 In this context, we present here our analysis of 169 transit light curves observed in the optical by the K2 \citep{Luger2017a}, SPECULOOS \citep[Gillon 2018]{Burdanov2017} and Liverpool \citep{Liverpool} telescopes. We combine our measurements with the ones obtained in the mid-IR by Spitzer/IRAC \citep{Delrez2018} and in the near-IR by HST/WFC3 \citep{deWit2018} to construct the broadband transmission spectra of the TRAPPIST-1 planets over the 0.8-4.5 $\mu$m spectral range. We confront these spectra with stellar contamination models in order to assess the impact of the heterogeneity of the star's photosphere on the atmospheric characterization of its planets. 

   The new observations and their reduction are described in Section \ref{dataanalysis}, as well as our detailed data analysis and results. In Section \ref{discussion} we discuss the temporal variability of the measured transit depths, as well as the structure of the planets' broadband transit transmission spectra, notably leveraging the visible part of these spectra for the first time.  We present two different scenarios able to fit the spectra, and for which stellar heterogeneity could be dominated by a few giant cold spots or a few small hot faculae, and discuss their implications for the atmospheric characterization of the planets. Finally, we give our conclusions in Section \ref{conclusion}.


\section{Observations and data analysis} \label{dataanalysis}
\subsection{Observations}
The new data used in this work consists of  transit light curves of the TRAPPIST-1 planets observed from the ground by the SPECULOOS \citep{Gillon2018} and Liverpool \citep{Liverpool} telescopes and from space by the K2 mission \citep{Howell2014}.

We observed 37 different transits with 1 or 2 telescopes of the SPECULOOS-South Observatory \citetext{SSO, \citealt{Burdanov2017}, \citealt{Gillon2018}} at Cerro Paranal, Chile (see Table 1), in the context of the commissioning of the facility. This represents 52 transits in total as some were observed with two SSO telescopes simultaneously. Each SSO robotic telescope has a primary aperture of 1m and a focal length of 8m, and is equipped with a 2k$\times$2k deep-depletion CCD camera whose 13.5 $\mu$m pixel size corresponds to 0.35" on the sky (field of view = 12$^\prime$x12$^\prime$). These observations were carried out in an I+z filter for which we computed an effective wavelength of $\sim$0.9$\mu$m for a M8-type star like TRAPPIST-1, taking into account the spectral response curve of the telescope+atmosphere. Exposure times of 23s were used for all observations. A standard calibration (bias, dark and flat-field corrections) was applied to each image, and fluxes were measured for the stars in the field with the DAOPHOT aperture photometry software \citep{Stetson1987}. Differential photometry was then performed after a careful selection of comparison stars.

We obtained 13 transits of the TRAPPIST-1 planets with the use of 2-m Liverpool Telescope \citep[LT, ][]{Liverpool} installed on the island of La Palma at the Roque de los Muchachos observatory. For our observations, we used the IO:O optical wide field camera which has 4k$\times$4k deep-depletion CCD with 15~$\mu$m-sized pixels and 10$\times$10~arcmin$^2$ field of view. We used 2$\times$2 binning what resulted in 0.3 arcsec~pixel$^{-1}$ image scale. All the observations were performed in Sloan z' band with 20 sec exposures. Data reduction and subsequent aperture photometry were carried out in the same manner as for the SSO data. 

TRAPPIST-1 was observed with the K2 telescope in an overall bandpass ranging from 420 to 900 nm over a period of 79 days in Campaign 12, which represents a total of 104 transits. The short cadence Target Pixel File (TPF), with a cadence rate of 1-per-minute, was downloaded from the Mikulski Archive for Space Telescope (MAST). We used the same procedure to extract and detrend the lightcurve as in \cite{Luger2017a} and \cite{Grimm2018}. We first applied a centroiding algorithm to find the (x,y) position of the PSF center in each cadence frame. We summed the flux within a circular top-hat aperture, centered on the PSF center in each frame. We used a Gaussian Process regression pipeline  (\cite{Luger2017a}, \cite{Grimm2018}) to remove the instrumental systematics due to K2 telescope's periodic roll angle drift, and the stellar variability. The systematics were fitted using a kernel that contained additive terms for the time- and position-dependent variation, enabling us to separate and subtract them individually. To ensure that the transits were not fitted as stellar variability, we masked them out during the fitting and regression procedure. The stellar and long-term variability was then subtracted from the light curve. The 6-hour combined differential photometric precision (CDPP) of the detrended lightcurve is 339 ppm.

We considered only well-isolated and complete transits in our analysis, discarding blended transits of different planets (9 transits discarded), partial transits (6 transits discarded), transits affected by flares (7 transits discarded), and transits affected by technical problems or bad weather conditions (3 transits discarded). In total 35 transits were discarded. Our final dataset was composed of 169 transit light curves, respectively 67 for TRAPPIST-1~b, 45 for -1~c, 21 for -1~d, 18 for -1~e, 8 for -1~f, 7 for -1~g, and 5 for -1~h. The number of transits kept for each planet is presented in Table \ref{nb_transit} for K2, SSO, and LT.

\begin{table}[h!]
\centering                          
\begin{tabular}{c c c c }        
\hline\hline                 
Planet  & K2 & SSO & LT  \\    
\hline                        
   TRAPPIST-1\,b  & 42  & 20 & 4 \\      
   TRAPPIST-1\,c  & 29  & 11 & 5\\  
   TRAPPIST-1\,d  & 15  &  5 & 1\\  
   TRAPPIST-1\,e  &  8  &  8 & 2\\  
   TRAPPIST-1\,f  &  6  &  2 & /\\  
   TRAPPIST-1\,g  &  3  &  3 & /\\  
   TRAPPIST-1\,h  &  1  &  3 & 1\\  

\hline                                   
\end{tabular}
\caption{Number of transits observed by K2, SSO, and LT analyzed in this work for each TRAPPIST-1 planet.}             
\label{nb_transit}
\end{table}

\subsection{Data analysis}

We chose to follow different approaches in our data analysis to ensure the robustness of our results. First, we analyzed each transit individually to extract their individual properties to, notably, search for signs of variability. Then, we proceeded to a global analysis of all transit light curves for each planet to determine precisely the average transit depths in K2, SSO, and LT bandpass. Finally, we performed an additional global analysis, this time enabling all transits to have different depths in order to assess their variability. For those two distinct global analyses, the transits observed by K2, SSO, and LT were analyzed separately. 
All of our analyses were performed with the most recent version of the adaptive Markov Chain Monte-Carlo (MCMC) code introduced in \cite{Gillon2012} \citep[see][hereafter G14, for an extensive description of our MCMC algorithm]{Gillon2014}. In this work we assumed a quadratic limb-darkening law for all the analyses, using normal prior distributions for the limb-darkening coefficients $u_{1}$ and $u_{2}$ based on theoretical values and 1$\sigma$ errors interpolated from the tables of \cite{Claret2011}. The modes of the normal prior distributions for $u_{1}$ and $u_{2}$ for the non-conventional I+z filter used by SSO were chosen as the average of the values interpolated from the tables for the standard filters $I_{c}$ and $z'$.  

Finally, for each instrument we also performed a global analysis of all transits for each planets with free limb-darkening (LD) coefficients, those values being the same across all planets within each global analysis. The aim of this analysis was to better constrain the limb darkening coefficients, as each planet samples a different chord of the stellar photosphere. For K2, the fitted limb-darkening coefficients through this procedure are consistent with the model-based limb-darkening priors used in the other analyses, the output LD coefficients from this global analysis were successfully constrained by the many transits. In this case, their respective values were: u1=1.00 +- 0.1 ;  u2=-0.04+-0.2 whereas the priors used on the LD coefficients in the rest of our analyses from interpolation of \cite{Claret2011} tables were u1=0.99 +- 0.09 ; u2=-0.19 +- 0.08, which is consistent. The  transit depths derived from this analysis are consistent with the remainder of our analyses (Appendix Table \ref{limb_darkening}). Unfortunately, for SSO and LT these global analyses failed to converge, meaning that the data do not allow for the constraint of the limb darkening coefficients.

\subsubsection{Individual analyses of the light curves}\label{individual}

First, we converted for each photometric measurement the mid-exposure time to the $BJD_{TDB}$ time system, as recommended by \cite{Eastman2010}. We modeled each transit with the model of  \cite{Mandel2002} multiplied by a baseline model accounting for the photometric variations of stellar, atmospheric, and instrumental origins (see G14). For each light curve, the model selection was based on the minimization of Bayesian Information Criterion \citep[BIC, ][]{Schwarz1978}. For a significant fraction of the light curves obtained by K2 and SSO, including a polynomial function of time in the model -to account for the low-frequency signals like the rotational variability of the star- resulted in a significant decrease of the BIC (see appendix Table.\ref{baseline_k2}). For some SSO and LT light curves, additional terms in the position or width of the stellar point-spread function were also favored (see appendix Table.\ref{baseline_spc}, Table. \ref{baseline_liverpool}). A small fraction of the SSO's light curves' baselines also included an airmass and/or a background polynomial function.

For each transit light curve, the jump parameters of the MCMC analysis, i.e. the parameters perturbed at each step of the Markov chains, were:

$\bullet$ The transit depth (planet-to-star area ratio) $dF=(R_{p}/R_{\star})^2$, the time of mid-transit (or inferior conjunction) $T_{0}$, and the transit impact parameter assuming a circular orbit $b$=$a\cos{i}/R_{\star}$, where $a$ is the semi major axis and $i$ the inclination of the orbit. 

$\bullet$ The mass, radius, effective temperature, and metallicity of the star, for which we assumed the following normal prior distributions: $M_{\star}=0.089 \pm 0.006 M_{\odot}$, $R_{\star}=0.121 \pm 0.003 R_{\odot}$, $T_{eff}=2516 \pm 41 K$, and $[Fe/H]=0.04 \pm 0.08$ \citep{VanGrootel2018}, respectively. 

We first assessed a Correction Factor $CF$ for each individual light curve via a short (10,000-steps) Markov chain. This  correction factor was then used to rescale the photometric error bars while accounting for a possible inadequate estimation of the white noise ($\beta{w}$) and the presence of red noise ($\beta{r}$) via $CF=\beta_{w}*\beta_{r}$. $\beta_{r}$ allows to account for possible correlated noise present in the light curve, this scaling factor is determined by following a procedure similar to the one described \cite{Winn2008} it is obtained by comparing the standard deviations of the binned and unbinned residuals for different binning intervals ranging from 5 to 120min, i.e. the typical time scales of an eclipse light curve (e.g. the duration of ingress or egress). 

. We then ran 2 chains of 100,000 steps for each light curve and successfully tested their convergence using the statistical test of \cite{Gelman1992}.

The results obtained from theses individual analyses are shown in Appendix Table \ref{individual_spc} for SSO, in Appendix Table \ref{individual_k2} for K2, and in Appendix Table \ref{individual_lt} for LT.
Each table gathers for each planet the transit times and depths derived from these individual analyses. The results are discussed in Section \ref{discussion}.

\subsubsection{Global analyses}\label{global}

Our next step was to perform, for each planet and for each dataset (K2, SSO, and LT), a global analysis of all transit light curves, to better separate the actual transit signals from the correlated noise of similar frequencies, and thus to improve the accuracies of the derived transit depths.	

These global analyses were done in two steps: first, for each planet and each instrument (K2, SSO, and LT), a general global analysis of all the transits with common transit shape parameters, followed by a global analysis allowing for transit depth variations.

We used the same priors on the stellar parameters as reported in Section. \ref{individual}. However, in this global analysis, we set a transit timing variation (TTV) as a jump parameter for each transit, fixing the planetary periods $P$ and reference transit timings $T_0$ to those reported in \cite{Delrez2018}. This global analysis includes 6 shared parameters across transits (the stellar parameters $M_{*}$, $T_{eff}$, $R_{*}$, [Fe/H] + limb darkening coefficients), for each planet the individual parameters are {\it df} and {\it b}, and same number of TTV than number of transits.

For each transit, we assumed the baseline model derived from the individual analysis, following the same procedure to rescale the photometric error bars, and derived our parameter estimates from the posterior distributions obtained from two Markov chains of 100,000 steps,  with 25\% burn-in phase, whose convergence was checked using the \cite{Gelman1992} test. The transit depths obtained for each data set are displayed in Table \ref{depth_glob}.

\begin{table}[h!]
\centering                          
\begin{tabular}{c c c c}        
\hline\hline                 
Planet  & $dF_{K2}$ (\%) & $dF_{SSO}$ (\%) & $dF_{LT}$(\%)\\    
\hline                        
   TRAPPIST-1\,b  & 0.721 $\pm$ 0.021 & 0.760 $\pm$ 0.025 & 0.746 $\pm$ 0.036 \\      
   TRAPPIST-1\,c  & 0.684 $\pm$ 0.019 & 0.736 $\pm$ 0.029 & 0.724 $\pm$ 0.027 \\
   TRAPPIST-1\,d  & 0.412 $\pm$ 0.028 & 0.354 $\pm$ 0.027 & 0.301  $\pm$ 0.071 \\
   TRAPPIST-1\,e  & 0.449 $\pm$ 0.034 & 0.453 $\pm$ 0.025 & 0.475 $\pm$ 0.054\\
   TRAPPIST-1\,f  & 0.541 $\pm$ 0.034 & 0.672 $\pm$ 0.052 & / \\ 
   TRAPPIST-1\,g  & 0.668 $\pm$ 0.070 & 0.755 $\pm$ 0.035 & /  \\ 
   TRAPPIST-1\,h  & 0.347 $\pm$ 0.058 & 0.321 $\pm$ 0.036 & 0.257 $\pm$ 0.035 \\ 
\hline                                   
\end{tabular}
\caption{Transit depths derived from the global analysis of all transits of each planet. Observations from K2, SSO, and LT were processed independently. }
\label{depth_glob}
\end{table}

In a second step, we thus performed similar global MCMC analyses, but this time with the depths of all individual  transits as jump parameters for all three instruments (K2, SSO, and LT). The aim here was to benefit from the constraint brought by the common transit shape (duration, impact parameter) to derive more accurate individual transit depths, and thus to better assess their potential variability. This time the analysis includes 4 shared parameters across transits (the stellar parameters $M_{*}$, $T_{eff}$, $R_{*}$, [Fe/H]), for each planet there is as many individual transit depths as transit plus the impact parameter (limb darkening coefficients are fixed) , and same number of TTV than number of transits. 

\begin{figure*}[h!]
\centering
\includegraphics[angle=0,width=7.5cm]{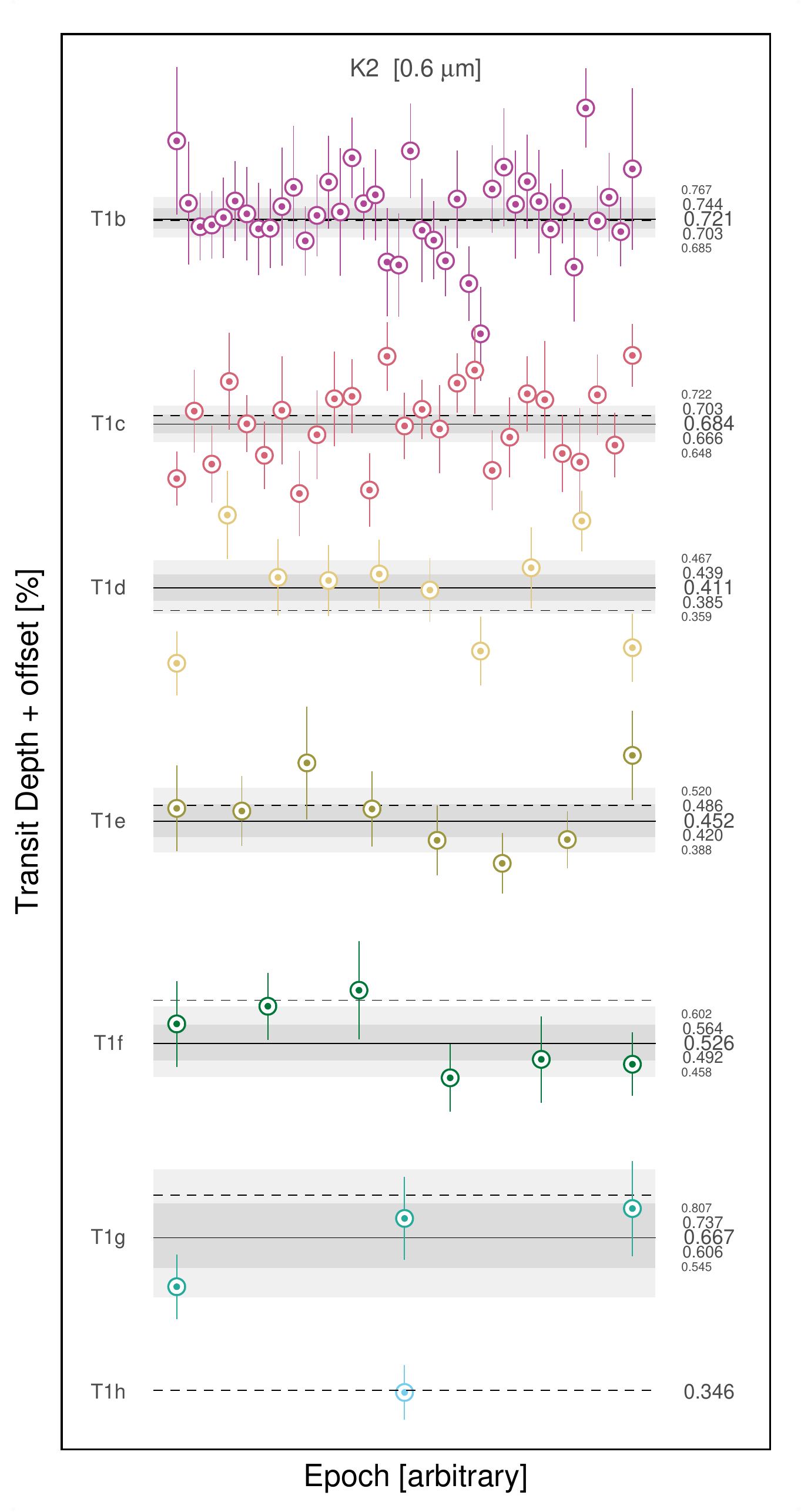}
\includegraphics[angle=0,width=7.5cm]{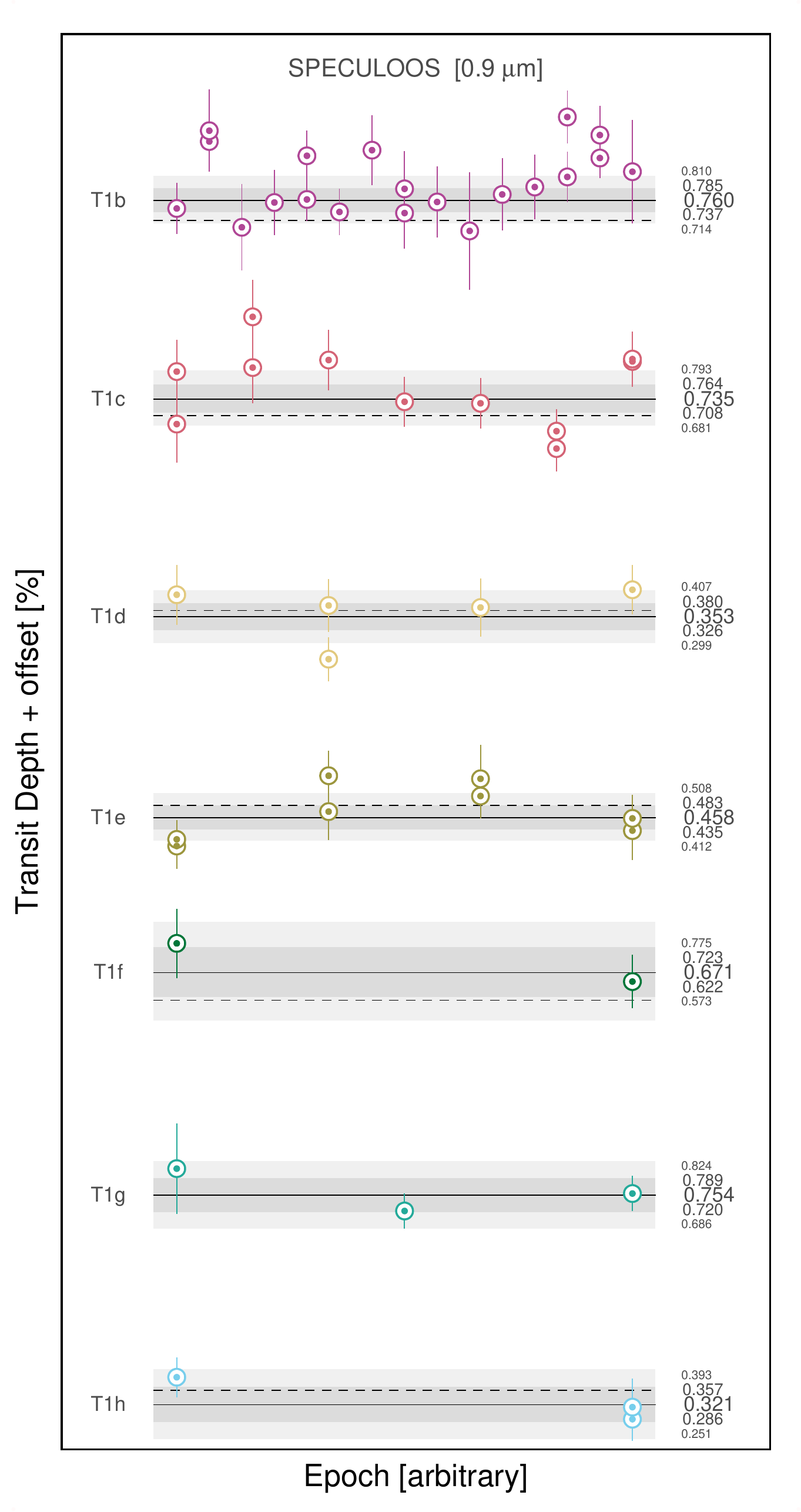}
\caption{{\it Left:} Evolution of the measured transit depths from the global analysis of transit light curves gathered by K2. The horizontal black lines show the medians of the global MCMC posteriors PDFs (with their 1 and 2$\sigma$ confidence intervals, in shades of grey), and dotted lines show the medians of the global MCMC posteriors PDFs for all transits of the same planet observed by {\it Spitzer}, as reported in \cite{Delrez2018}. Events are ranked in order of capture, left to right (but not linearly in time). {\it Right:} Similarly, but for transit observed with SSO. Neither SSO or K2 data show significant variability (less than 3$\sigma$).}
\label{depth_var_global}
\end{figure*}

Table \ref{global_spc}, \ref{global_k2} and \ref{global_liverpool} in the appendix present our measured transit depths as deduced from our global analyses of SSO, K2, and LT transits, respectively. Their temporal evolution is shown for each planet in Fig. 
\ref{depth_var_global} (we did not plot Liverpool data because of the few number of light curves, but the values can be found in Table \ref{global_liverpool}). For further comparison, these figures also display the medians of the global MCMC posterior probability distribution functions (PDFs)  as measured with {\it Spitzer} at 4.5 $\mu$m by \cite{Delrez2018}, and also the PDF derived from the MCMC analyses assuming common transit depths. 

We compared the results obtained from the individual and global analyses of the transits and found them to be fully consistent. Accurately constraining the transit shape through a global analysis slightly improves the errors on the depths or timings for some transits, while others have larger errors due to the clearer separation between signal and red noise. For this reason, we adopt the results of our global analyses as our final ones.

\section{Results and Discussion}\label{discussion}

\subsection{Temporal evolution of the transit depths}\label{results}

Changes in the transit depths measured for a planet in a given bandpass could result from the evolution of stellar heterogeneities on or outside the chord transited by the planet. Fig. \ref{depth_var_global} shows the evolution of the transit depths derived from our global analyses of K2 and SSO light curves. These analyses assumed  a common transit profile -except for the depths- for each planet and each instrument to better separate the correlated noise from the transit signals and thus guarantee robust results on the transit depths. From those results, we notice that for all planets the depths are consistent from a transit to another, with no discrepancy larger than 3$\sigma$. We computed the standard deviation of the measurements and compare it to the mean value of the measurement errors for each dataset, the values are presented in Table. \ref{standard_dev}. 
\begin{table}[h!]
\centering                          
\begin{tabular}{c c c c c }        
\hline\hline                 
Telescope & Planet & \# transits & $\sigma$  & Mean error \\    
 & & & (\%) & (\%) \\
 \hline K2 & -1b  & 40& 0.084 & 0.14 \\ 
 & -1c & 27 & 0.080 & 0.081 \\ 
 & -1d & 10 & 0.11 & 0.073 \\ 
 & -1e & 8 & 0.077 & 0.080 \\ 
 & -1f & 6 & 0.072 & 0.080 \\
 & -1g & 3 & 0.087 & 0.085 \\
 & -1h & 1 & / &  / \\
\hline                        
SPECULOOS & -1b & 20 &0.069 & 0.067 \\ 
 & -1c & 11 &0.080 & 0.059 \\ 
 & -1d & 5 &0.057 & 0.053 \\ 
 & -1e & 8 &0.055 & 0.053\\ 
 & -1f & 2 &0.055 & 0.063 \\
 & -1g & 3 &0.044 & 0.055 \\
 & -1h & 3 &0.044 & 0.047 \\
 \hline
 Liverpool &  -1b & 3 &0.087 & 0.081 \\
 & -1c & 4 &0.102 & 0.062 \\
 & -1e & 2 &0.087 & 0.081 \\
\hline                                   
\end{tabular}
\caption{Standard deviation and mean errors of the measured transit depths   for all data set. {\it Remark:} there are no values for planet h with K2 nor planets d, g, h with the Liverpool telescope because we had only one light curve for each of those planets.}             
\label{standard_dev}
\end{table}

We found that the standard deviation is consistent with the mean of the measurements errors for most of the planets/instruments associations. The exceptions are planet c (SSO, LT) and planet d (K2), where the dispersion of the measurements is actually larger than the mean errors. These mild discrepancies could be genuine, but they could also  originate from small-number statistics. Indeed, only 4 transits are used to compute the statistics for LT, 11 transits for SSO for planet c, and 10 transits for planet d.

Looking at the few transits that were observed simultaneously with {\it Spitzer} (values from \citep{Delrez2018}) and K2 (see Table. \ref{individual_k2}) on one hand and with SPECULOOS  (see Table. \ref{individual_spc}) and LT (see Table. \ref{individual_lt}) on the other hand, we see that the transit depths values are in agreement with one another (see Table. \ref{shared_transits}), K2 error bars being significantly larger than {\it Spitzer} error bars. For certain transits, the value derived from K2 is larger than the one derived from {\it Spitzer}, while for others it is the opposite. We can conclude on the transit observed simultaneously by SPECULOOS and Liverpool as it is unique. 

\begin{table}[h!]
\centering                          
\begin{tabular}{c c c c }        
\hline\hline                 
Planet  & Epoch &  K2 & \it{Spitzer}   \\    
\hline                        
-1b & 318 & 0.830 $\pm$ 0.120 & 0.751 $\pm$ 0.027 \\
		& 320 & 0.669 $\pm$ 0.160 & 0.699 $\pm$ 0.023 \\
		& 321 & 0.988 $\pm$ 0.120 & 0.801 $\pm$ 0.028 \\
		& 325 & 0.866 $\pm$ 0.130 & 0.732 $\pm$  0.022 \\
		& 326 & 0.693 $\pm$ 0.073 & 0.724 $\pm$ 0.023 \\
		& 327 & 0.851 $\pm$ 0.086 & 0.663 $\pm$ 0.021 \\
        \hline                        
-1c  & 215 & 0.604 $\pm$ 0.090 & 0.672 $\pm$ 0.025 \\
		& 216 & 0.686 $\pm$ 0.080 & 0.652 $\pm$ 0.020 \\
        & 217 & 0.797 $\pm$ 0.120 & 0.735 $\pm$ 0.035 \\
        & 218 & 0.809 $\pm$ 0.400 & 0.674 $\pm$ 0.029 \\
        & 219 & 0.663 $\pm$ 0.071 & 0.668 $\pm$ 0.024 \\
        & 220 & 0.830 $\pm$ 0.120 & 0.725 $\pm$ 0.024 \\
                \hline                        
-1d  & 34 & 0.304 $\pm$ 0.130 & 0.384 $\pm$ 0.020 \\
		& 35 & 0.412 $\pm$ 0.210 & 0.382 $\pm$ 0.024 \\
        & 36 & 0.361 $\pm$ 0.110 & 0.348$\pm$ 0.019 \\
                \hline                        
-1f  & 15 & 0.494 $\pm$ 0.090 & 0.648 $\pm$ 0.025 \\
\hline
-1g  & 12 & 0.867 $\pm$ 0.170 & 0.777 $\pm$ 0.020 \\
&&&\\
\hline\hline                 
Planet  & Epoch &  SPECULOOS & Liverpool   \\    
\hline    
-1e  & 53 &  \makecell{0.522 $\pm$ 0.055 \\ 0.590 $\pm$ 0.057} & 0.476 $\pm$ 0.069 \\
\hline                              
-1h  & 17 &  \makecell{0.316 $\pm$ 0.057 \\ 0.291 $\pm$ 0.044} & 0.257 $\pm$ 0.035 \\
\hline
\end{tabular}
\caption{{\it Up:} Depth of transits observed simultaneously by K2 and {\it Spitzer}. {\it Down:} Same but for SPECULOOS and Liverpool telescope. }             
\label{shared_transits}
\end{table}

\subsection{Transmission spectra of the TRAPPIST-1 planets}\label{Transmission_spectra}

\begin{figure}[h!]

\centering
\includegraphics[angle=0,width=8cm]{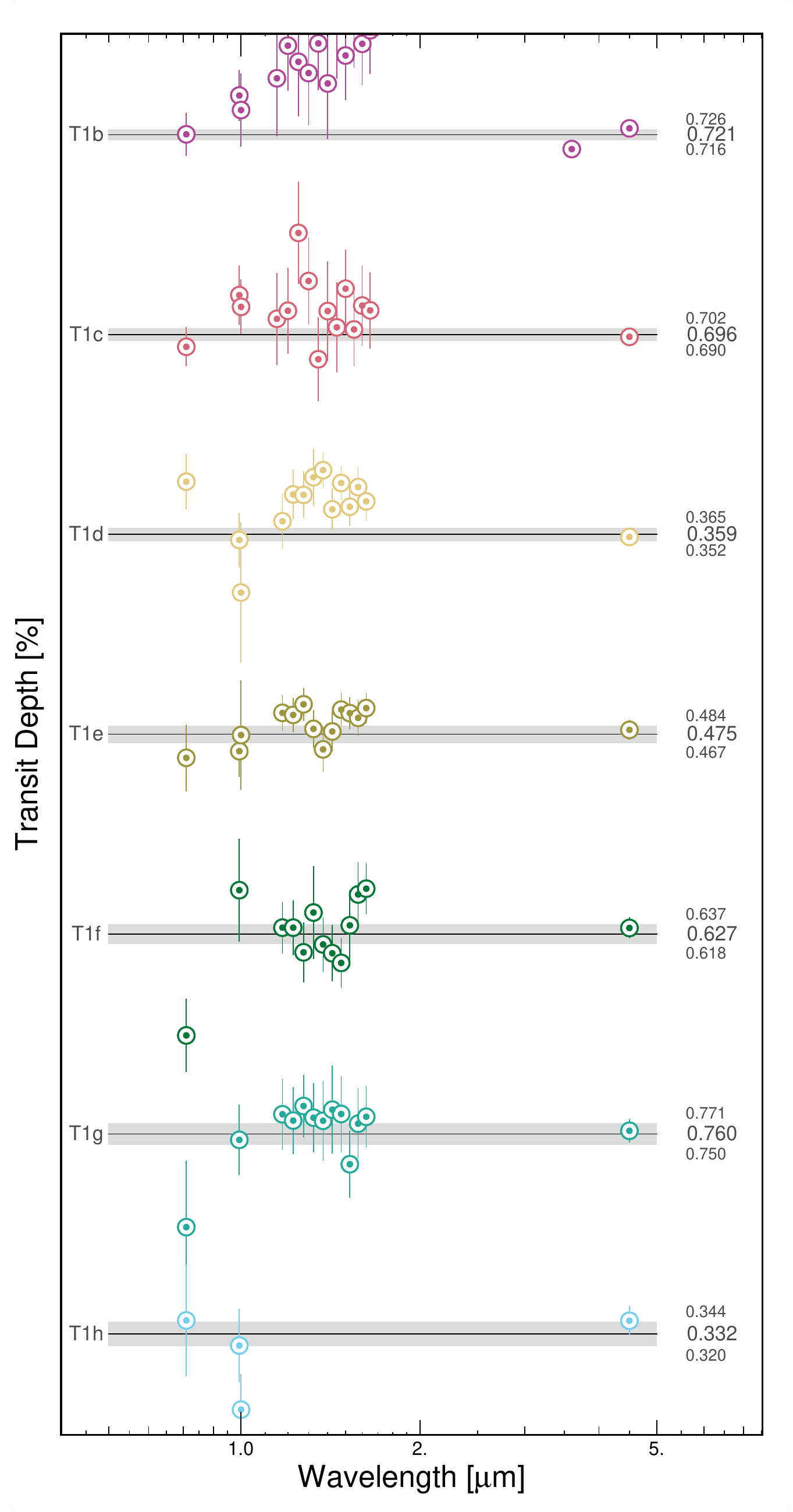}
\caption{Spectra of the seven TRAPPIST-1 planets. The continuous line is the weighted mean of all non-HST measurements for each planet (with its 1 $\sigma$ confidence, in shades of grey). Each point stands for the median of the global MCMC posterior PDF with error bars at the effective wavelength of the instrument (13 points (14 for T1b) per planet: one for K2, one for SSO, one for LT, 9 for HST/WFC3 and one (two for T1b, 3.6$\mu m$ and 4,5 $\mu m$) for $Spitzer$).}
 \color{black}
\label{spectra}
\end{figure}

Combining the results of our analyses to the ones presented by \citet{Delrez2018} for {\it Spitzer} measurements and by \citet{deWit2018} for HST/WFC3 measurements, we construct the broadband 0.8-4.5 $\mu$m transit transmission spectra of TRAPPIST-1 planets (Fig. \ref{spectra}).

We first note that although the measurements obtained with the HST data do not show features over the WCF3 band (1.1 to 1.7 $\mu$m), the transit depths are significantly deeper than those obtained at other wavelengths for planets b and d. Although this is intriguing, these deeper transits could very well have an instrumental origin. Indeed, as HST is on a low-Earth orbit, it can monitor TRAPPIST-1 for an average of $\sim$50 minutes per orbit out of the $\sim$95 minute orbital duration. The observation of a transit during an HST visit is typically based on 4 or 5 orbits. Due to the small transit durations of the TRAPPIST-1 planets, only one window per visit covers a transit. Yet, although the transit durations of TRAPPIST-1 planets are short, they have roughly the same duration of HST's observation window leading to a small (and at times negligible) constraint on the baseline level from the in-transit orbit. As HST/WFC3 spectrophotometric observations are affected by orbit-dependent systematic effects, such a limited constraint on the baseline level from the orbit constraining the transit depth can result in a diluted or amplified monochromatic transit depth. The current measurements are particularly limited in such joint ``transit depth--baseline level'' measurements for planet b \citep[see Fig. 1 of][]{deWit2016} and planet d \citep[see Fig. 1 of][]{deWit2018}--and reduced for planets c and e--which is consistent with the level of discrepancies seen in Fig. \ref{spectra}.
%
%
We also note that the transit depth measured for planet f at 0.6$\mu$m (K2) is  $\sim$3-$\sigma$ shallower than the mean of the other measurements. This measurement could be explained by its low statistical significance (only 6 transits) or by the detrending of K2 systematic effects and significant stellar variability applied to the light curve before its modeling (see Section 2.1). Nevertheless, there seems to be no significant biases from detrending in the other planets measurements so we would better wait for the analyses of additional transits of planet f in this bandpass to confirm or discard this value. For the other planets, no significant chromatic variation is observed. We note that an argument against a stellar contamination origin of the structure visible in the transit spectra of planets b, d, and f, is the absence of similar structures for planets with similar transit impact parameters, i.e. transiting nearly the same chords of the stellar disk. 

\begin{figure*}[h!]

\centering
\includegraphics[angle=0,width=8.8cm]{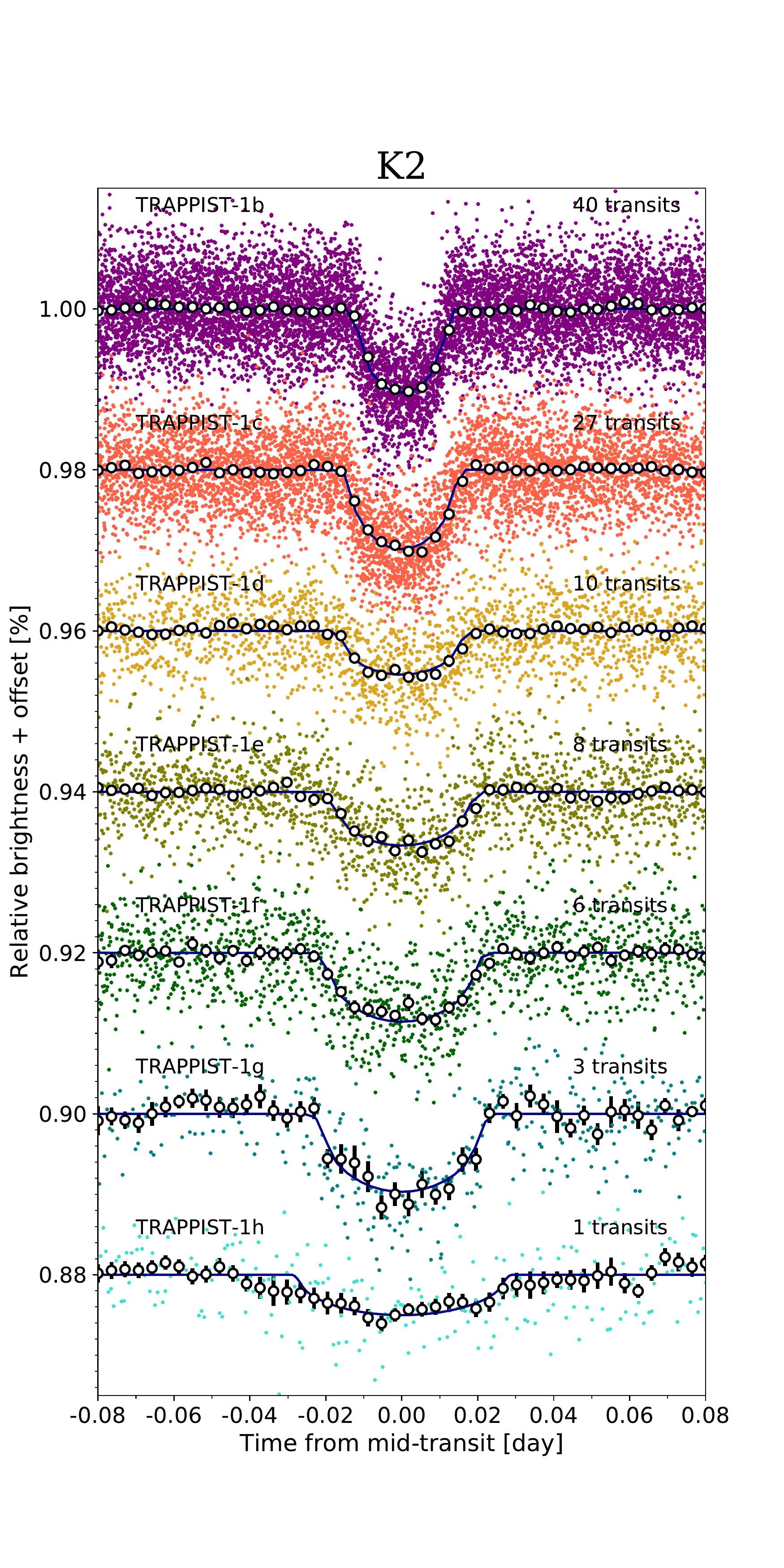}
\includegraphics[angle=0,width=8.8cm]{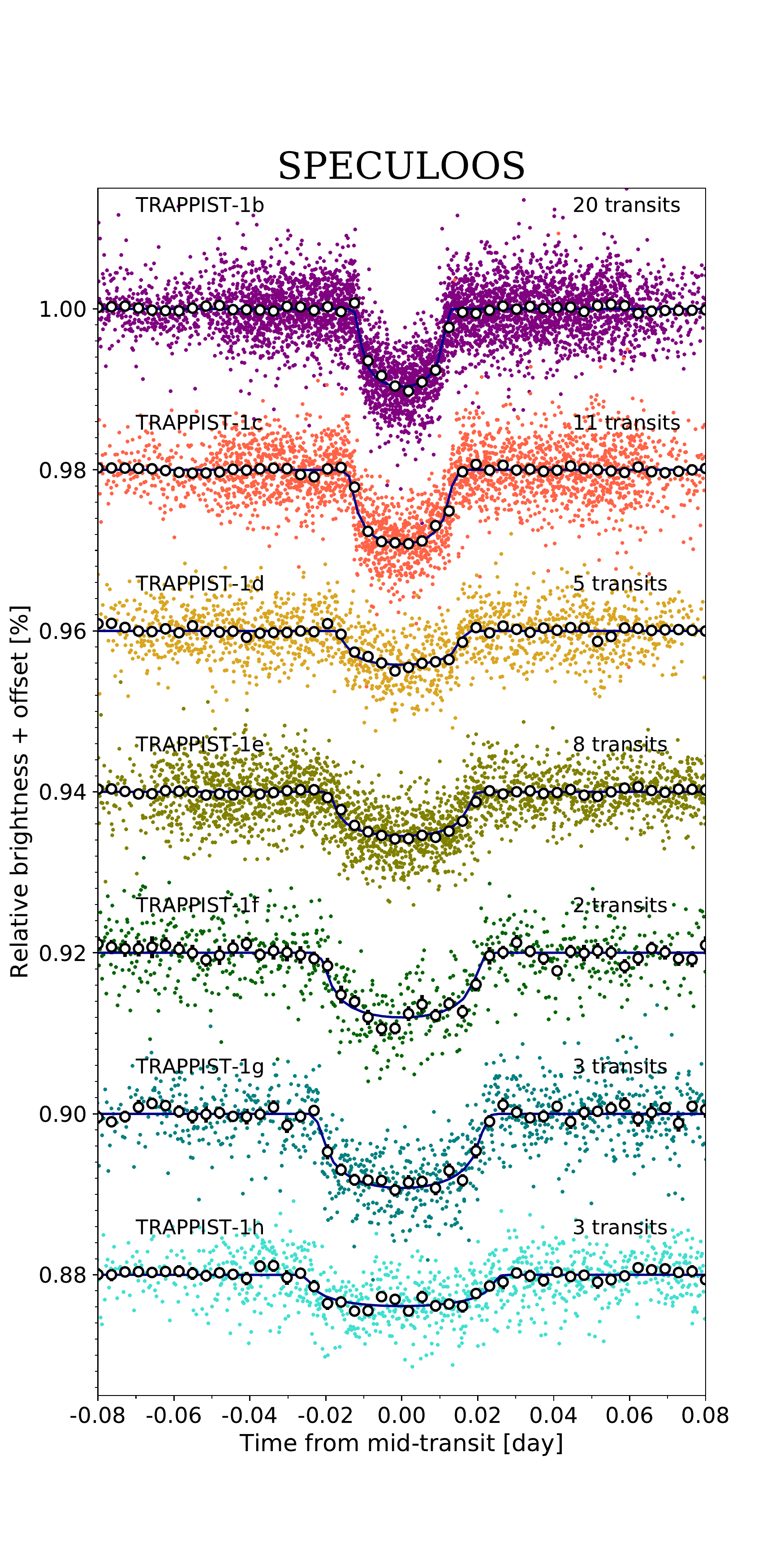}
\caption{{\it Left:} Period-folded photometric measurements obtained by K2 near the transits of the seven planets, corrected for the measured TTVs. Colored dots show the unbinned measurements; open circles depict the 5minute-binned measurements for visual clarity. The best-fit transit models are shown as dark blue lines. The numbers of transits that were observed to produce these combined curves are written on the plot. {\it Right:} Similarly but for SSO. }
\label{period_folded}
\end{figure*}

Fig. \ref{period_folded} shows the detrended period-folded photometry measured for each planet by K2 and SPECULOOS, as well as the corresponding best-fit transit model. A visual inspection of all individual transit light curves did not reveal such crossing events neither. \\

\subsection{Confrontation with the stellar contamination model of Z18}

The strong stellar contamination inferred for TRAPPIST-1 planets by Z18 is based on the  model presented by \cite{Rackham2017a}, which assumes an heterogeneous photosphere composed of unocculted spots and faculae, and is described by the equation:

\begin{equation}
\epsilon_{\lambda,s+f}=\frac{1}{1-f_{spot}(1-\frac{F_{\lambda,spot}}{F_{\lambda,phot}})-f_{fac}(1-\frac{F_{\lambda,fac}}{F_{\lambda,phot}})},
\label{epsilon_R18}
\end{equation}
in which $\epsilon_{\lambda,s+f}$ is the ratio of the observed transit depth $D_{\lambda ,obs}$ by the nominal transit depth $D_{\lambda}$ (i.e., the square of the true wavelength-dependent planet-to-star radius ratio) and represents the stellar contamination at wavelength $\lambda$; $F_{\lambda,phot}$, $F_{\lambda,spot}$ and $F_{\lambda,fac}$ refer to the flux of the mean photosphere, spots and faculae respectively; and $f_{spot}$ and $f_{fac}$ refer to the unocculted spot- and faculae- covering fractions \citep{Rackham2018}.

The contamination spectrum $\epsilon_{\lambda,s+f}$ was then multiplied with an assumed wavelength-independent nominal planetary transit depth by Z18 to obtain a transit spectrum whose wavelength-dependence is only due to the stellar contamination. Ultimately, they fitted the percentages of spots and faculae covering fractions, as well as their temperatures and that of the mean photosphere, to represent at best the transit spectra of the TRAPPIST-1 planets that they measured from the HST/WFC3 presented in \cite{deWit2016} and \cite{deWit2018}. The authors chose to combine spectra of several planets, justifying their choice by the improved signal-to-noise ratio in detecting common
spectral features. To enable a straightforward comparison with the Z18 results, we added our measured transit depths of different planets to obtain the same combinations used by Z18.

The transit depth values obtained from our global analysis of K2, SSO, and LT transits, plus the values measured at 4.5 $\mu$m with {\it Spitzer} by \cite{Delrez2018}, and at 1.1-1.7 $\mu$m with HST/WFC3 by \cite{deWit2016} are displayed in Fig.\ref{Zhang_vs_obs} for the combination of planets b and c and Fig.\ref{Zhang_vs_obs_bcdefg} in Appendix for b+c+d+e+f+g, superimposed with the best-fit stellar contamination model of Z18. Appendix Table \ref{depth_value_combined} gathers the results for those two combination as well as the other combination used in Z18 (d+e+f+g). 

\begin{figure}[h!]
\centering
\includegraphics[angle=0,width=12cm]{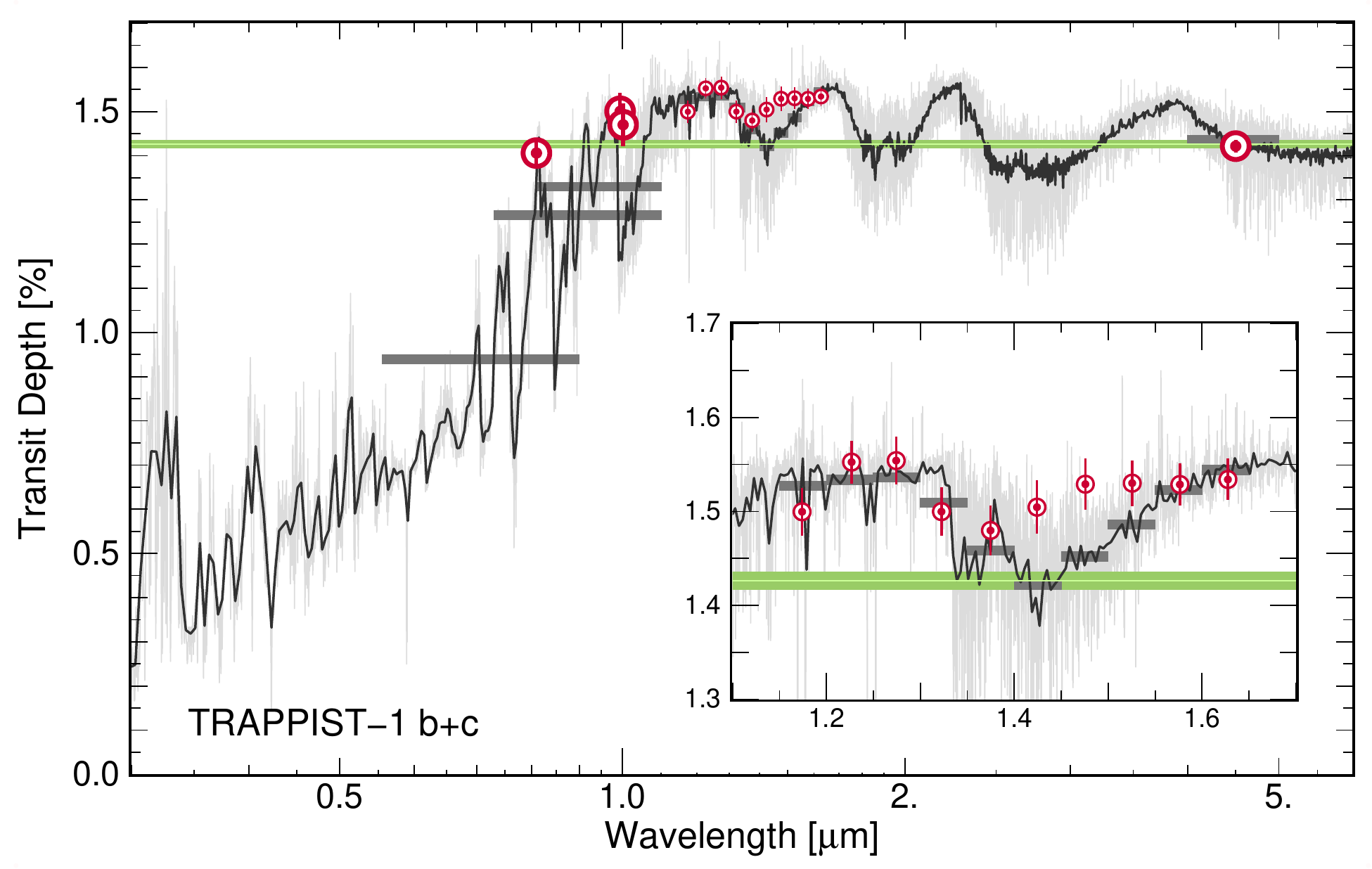}
\caption{Up: Comparison of the stellar contamination spectrum inferred by Z18 for TRAPPIST-1\,b+c transits [\cite{Zhang2018}] at two different resolutions (continuous black line and gray line) with the K2, SSO and LT measurements presented in this work, and the {\it Spitzer} and HST/WFC3 presented in \cite{Delrez2018} and \cite{deWit2016}, respectively (red points). The green line represents the weighted mean of all measurements except HST for the reasons outlined earlier in Section. \ref{Transmission_spectra}. Finally, the gray horizontal bars are the band-integrated value for the Z18 model on the effective bandpass of each filter (define as the interval were the product of the filter response and the stellar spectrum is greater than 1\%).
}
\color{black}
\label{Zhang_vs_obs}
\end{figure}

The expected transit depths from the best-fit stellar contamination model of Z18, integrated over the spectral bands of the observations, are reported in Appendix Table \ref{depth_value_combined} for the combination of planets b+c, b+c+d+e+f+g, and d+e+f+g, along with the actual measurements. To compute those values, we multiplied the contamination spectrum $\epsilon_{\lambda,s+f}$ inferred in Z18 by the maximum combined transit depth for the corresponding combination of planets $D_{b+c}$, measured from HST/WFC3 data by \cite{deWit2016}.
%

As shown in Fig. \ref{Zhang_vs_obs} and Table \ref{depth_value_combined}, the  dramatic drop of the transit depth in the visible predicted by Z18 model is not observed. As a matter of fact, the Z18 prediction for K2 bandpass are discrepant by more than $10\sigma$ from the observations, at $\sim 6.5\sigma$ for SSO, at $\sim 3.5\sigma$ for {\it Liverpool}, and $\sim 1.4\sigma$ for {\it Spitzer}. The contamination model inferred by Z18 can thus be firmly discarded. It should also be noted that Z18 attributed an inverted water absorption spectral feature to low-significance variations present in their analysis of the HST measurements. However, in \cite{deWit2016} data we do not see significant traces of this inverted water absorption feature (see zoomed box in Fig. \ref{Zhang_vs_obs}).

Finally, in Z18, the sum of the spot and faculae covering fraction approaches 100\% with spot of size $R_{spot}= (1.63 \pm 0.50) \times 10^{3} km$ \citep{Rackham2017a}, while we know from \citet{Delrez2018} that the chords of transit of the TRAPPIST planets cover at least 56\% of a stellar hemisphere. Z18's model should therefore predict a significant number of spot crossing event with amplitudes of the order of 400ppm \citep{Rackham2017a}. Quantitatively, according to Z18 for T-1b+T-1c we would expect a frequency rate of 18\% spot crossing and 34\% of faculae crossing events. We analysed all light curves individually, we see comparable variability in and out of transit, at a significantly lower level than expected (maximum 200ppm) and no asymmetries in the amplitude of the residuals.

While the model of Z18 is discarded by our data, a significant stellar contamination of TRAPPIST-1 planets' transmission spectra remains a possibility. Indeed, the star's photosphere is definitely heterogeneous, as its K2 photometry shows a quasi-periodic variability of a couple \% with a dominant period of 3.3d that is consistent with the rotation of an evolving inhomogeneous photosphere \citep{Luger2017a}, or with the characteristic timescale between flares followed by spot brightening \citep[][hereafter M18]{Morris2018}. The photometry of the TRAPPIST telescope \citep{TRAPPIST} also shows variability of similar amplitude, with a dominant period identified to be $\sim$1.4d by \cite{Gillon2016}. We note that this latter value is close to the alias of 3.3d, suggesting that the periodogram analysis done by \cite{Gillon2016} did not identify the right period because of the discontinuous sampling of the TRAPPIST observations, or that the variability is only quasi-periodic.


\subsection{On the possible photospheric structure of TRAPPIST-1}
\subsubsection{Giant cold spots?}
While not stated explicitly, the photospheric model of Z18 considered solar-like spots + faculae, and not giant spots + faculae, as this is the only way for the percentages obtained for the best fit ($\sim 30\%$ of spots and $\sim 63\%$ of faculae) to agree to a certain extent with the predictions of R18 on which it is based ($8_{-7}^{+18}\%$ of spots and $54_{-46}^{+16}\%$ of faculae).  
At this point, it is worth explaining what is meant by giant spots and solar spots. The ``solar spot'' model used in R18 relies on small time-steady rotating spots to produce the predicted variability amplitude in transit depth. As the variations in flux cancel out when the spots rotate onto and off of the visible photosphere, a large number of spots are required to reach the predicted transit depth variation, leading to a large, heterogeneous, but nearly time-steady component. Conversely, the ``giant spot'' model shows large amplitude variability with small covering fraction as there is no cancellation between spots rotating on and off, and giant spots therefore have a variable component.

If instead of considering solar-type spots + faculae, we consider giants spots + faculae, we notice that the prediction from CPAT (composite photosphere and atmospheric transmission) model of \cite{Rackham2017a} on the transit depth variations are much less pessimistic (not more than 0.7\% difference between transit depth at $4.5\mu m$ and at $0.6\mu m$ for an M9V type star, R18, Fig. 7). We could thus imagine that the photosphere of TRAPPIST-1 is more likely to host giant spots than solar-like spots. In this case it is worth noticing that according to the predictions of R18, for Earth-twin type planets, the stellar heterogeneity does not jeopardize the detection of planetary atmospheric features with JWST anymore. Considering a precision of 30ppm with JWST, R18 indicates that for a M8V type star like TRAPPIST-1 the depth variations due to atmospheric features should be of the order of 90ppm whereas the variations due to stellar heterogeneity should be of the order of $\approx$17ppm, consequently allowing detections of planetary features despite stellar contamination.

As discussed above, the TRAPPIST-1 planets cover a significant part of the hemisphere of the star from latitudes up to $30^{\circ}$, latitudes where we find spots on the Sun \citep{Miletskii2009}. The next logical step is to look for giant spot-crossing events in the transits of the TRAPPIST-1 planets. In the observations carried out by {\it Spitzer} the in and out of transit variability was more likely attributed to systematic effects or granulation variability  \citep[see][]{Delrez2018}. Yet the spot-to-photosphere contrast is wavelength-dependent such that spot-crossing events are not detectable at all wavelengths \citep[see ][]{Ballerini2012}. However, our analyses of observations in the visible and near-IR carried out by K2, SPECULOOS and Liverpool telescope do not show transit depth variability that could have been attributed to stellar spot crossings during transits (see Section. \ref{results}). A possible scenario allows for giant spots consists of high-latitude spots that never cross the planets' transit chords, in a similar manner as the circumpolar spots observed for young mid- to late-type M-dwarfs not older than 1 Gyr \citep[see ][]{Barnes2015}; this potentially could explain the variability detected in the K2 bandpass. However, TRAPPIST-1 is not a young dwarf, its age having been estimated to be 7.6 $\pm$ 2.2 Gyr by \cite{Burgasser2017}, and the out-of-transit rotational variability resulting from a giant, dark polar spot does not match the small observed variability of 2ppm \citep{Delrez2018} seen in the infrared \citep{Morris2018}.  In addition, the giant spot model is disfavored by the correlations between flares and spot brightening seen in the K2 dataset, which indicates that the brightening is not due to spots rotating out of view, but rather due to a temporary brightening of the star which follows each flare event \citep{Morris2018}.
%
%
%
%
%

\subsubsection{Small hot faculae?}

In their studies, R18 and Z18 assumed that the active regions of TRAPPIST-1 are qualitatively similar to solar active regions in the spot and facular flux contrasts, and in the relative areas of each component. However, there is abundant evidence that the Sun is a poor analog for the starspot distributions of fully-convective stars \citep{Donati2003,Morin2008,Morin2010,Barnes2015}, which are likely driven by a different magnetic dynamo process \citep{Donati2011,Reiners2012}. 

\citet{Morris2018} presented an alternative, empirically-driven hypothetical spot distribution for TRAPPIST-1, consisting of a few small, bright (hot) spots. The proposed hot spots, which are correlated with the brightest flares, drive the modulation with an 3.3 day period in the K2 bandpass without generating a corresponding signal in the {\it Spitzer} 4.5 $\mu$m band, in agreement with the observations. 

\begin{figure}[h!]
\centering
\includegraphics[angle=0,width=9cm]{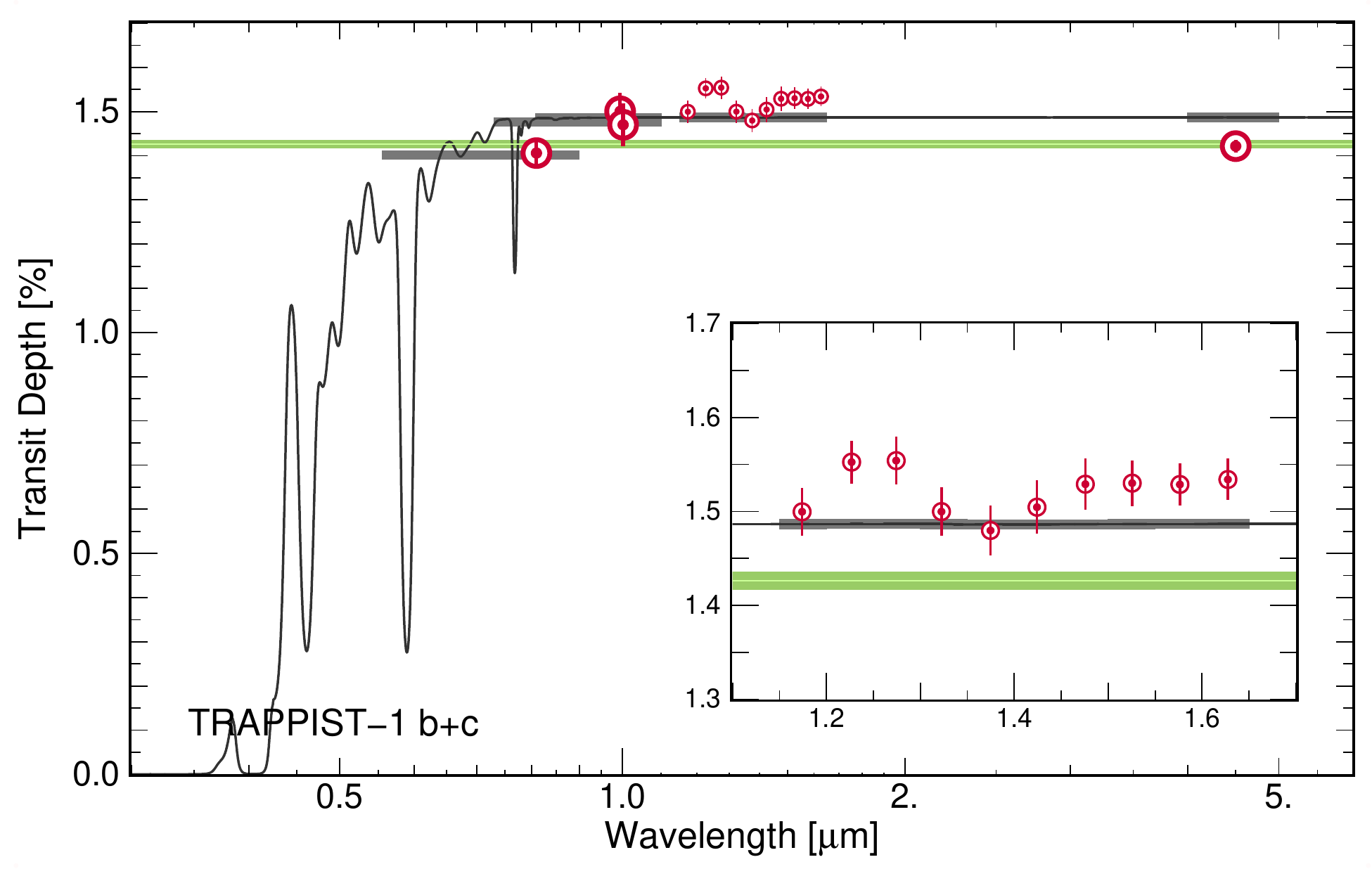}
\caption{Comparison of the observed transit depth variation (red points)  with the predictions from stellar contamination due to the bright spots proposed by \citet{Morris2018} for spots at 4500K (gray continuous line). We used PHOENIX model atmospheres with photospheric temperature 2511 K and the hot spot properties in M18.}
\color{black}
\label{brett4500K}
\end{figure}

We predict the effect of the hot spots of \citet{Morris2018} at 4500 K on the transit depths of TRAPPIST-1 b and c in Fig. \ref{brett4500K}. These spots produce a nearly-flat contamination spectrum for wavelengths $\gtrsim 0.7 \mu$m, and modest flux dilution (shallower transit depths) in the K2 bandpass. We find that spots with temperatures up to 4500 K are consistent at $\sim 2 \sigma$ with the observed transit depths, excluding the HST data for the reasons discussed above.

\section{Conclusion} \label{conclusion}


We performed individual and global analyses of 169 transit light curves obtained from space with K2 and from the ground with SSO and LT as well as the light curves obtained from mid-IR observations with {\it Spitzer} and near-IR with HST/WFC3 to construct the broadband transmission spectra of the TRAPPIST-1 planets over the 0.8-4.5$\mu m$ spectral range. While we could not find any significant temporal variability of the transit depths measured by the same instrument, our analysis reveals chromatic structures at the level of only 200-300ppm in the transit transmission spectra of planets b, d, and f. These results enable us to discard the highly heterogeneous photospheric model presented by Z18 and their subsequent conclusions regarding the potential of JWST to characterize the atmospheric properties of TRAPPIST-1 planets by transit transmission spectroscopy. We identify two possible photospheric structures for TRAPPIST-1 that could agree with our results, one dominated by a few high-latitude giant (cold) spots,  which is disfavored for different reasons, and the other by a few small and hot ($>$ 4000K) faculae. Although our measurements do not confirm the conclusions of Z18, they cannot rule out a significant stellar contamination of the planets' transmission spectra. The recent announcement of the delayed launch of JWST gives us the opportunity to investigate further the photospheric structure of TRAPPIST-1 -notably through photometric monitoring at different wavelengths- and its impact on the planets' transmission spectra. Furthermore, the JWST delay offers more time for the development of new strategies to optimally disentangle the stellar (contamination) and planetary (transmission) effects.

\section{Acknowledgement}

We thank Jon Marchant and Chris Copperwheat for their kind and frequent help in scheduling the Liverpool Telescope. The Liverpool Telescope is operated on the island of La Palma by Liverpool John Moores University in the Spanish Observatorio del Roque de los Muchachos of the Instituto de Astrofisica de Canarias with financial support from the UK Science and Technology Facilities Council.

The research leading to these results has received funding from the European Research Council (ERC) under the FP/2007-2013 ERC grant agreement no. 336480, and under the H2020 ERC grant agreement no. 679030; and from an Actions de Recherche Concert\'ee (ARC)
grant, financed by the Wallonia-Brussels Federation. This work was also partially supported by a grant from the Simons Foundation (PI Queloz, grant number 327127), as well as by the MERAC foundation (PI Triaud). LD acknowledges support from the Gruber Foundation Fellowship. VVG and MG are F.R.S.-FNRS Research Associates. JdW is grateful for the financial support received for the SPECULOOS Project from the Heising-Simons Foundation, P. Gilman, and C. \& L. Masson. EJ is F.R.S.-FNRS.  EA acknowledges USA NSF grant 1615315, the Guggenheim Foundation, and NASA Virtual Planetary Laboratory.
Senior Research Associate. B-OD acknowledges support from the Swiss National Science Foundation in the form of a SNSF Professorship (PP00P2\_163967). AJB acknowledges funding support from the US-UK Fulbright Scholarship programme.

\newpage
\appendix 



\label{AppendixA} 
\section{Description of the data}

\begin{center}
\begin{longtable}{|l|l|l|l|l|l|l|l|l|l|}
\caption{Description of transit light curves measured for TRAPPIST-1 planets by SPECULOOS-South. } 
\label{baseline_spc} \\
\hline
\hline \multicolumn{1}{|c|}{\textbf{Planet}} & \multicolumn{1}{|c|}{\textbf{Date}} & \multicolumn{1}{|c|}{\textbf{Telescope}} & \multicolumn{1}{|c|}{\textbf{Number of points}} & \multicolumn{1}{c|}{\textbf{Epoch}} & \multicolumn{1}{c|}{\textbf{Baseline}} & \multicolumn{1}{|c|}{\textbf{$\beta_{w}$}} & \multicolumn{1}{|c|}{\textbf{$\beta_{r}$}} & \multicolumn{1}{|c|}{\textbf{CF}} \\ \hline 
\endfirsthead

\multicolumn{3}{c}%
{{\bfseries \tablename\ \thetable{} -- continued from previous page}} \\
\\ \hline 
\endhead

\hline \multicolumn{3}{|r|}{{Continued on next page}} \\ \hline
\endfoot

\hline \hline
\endlastfoot
\hline
b	& 18 Jun. 2017	&	Europa	&	487	&	398	&	$p(fwhm^{1})$	&	1.22	&	1.20	&	1.47	\\
	&	30 Jun. 2017	&	Io	&	196	&	406	&	$p(t^{1})$	&	1.04	&	1.00	&	1.04	\\
	&	30 Jun. 2017	&	Europa	&	242	&	406	&	$p(t^{1})+p(xy^{1})$	&	1.02	&	1.89	&	1.93	\\
	&	01 Aug. 2017	&	Europa	&	273	&	427	&	$p(fwhm^{1})$	&	1.28	&	1.09	&	1.40	\\
	&	07 Aug. 2017	&	Europa	&	228	&	431	&	$p(fwhm^{1})$	&	1.07	&	1.49	&	1.59	\\
	&	13 Aug. 2017	&	Europa	&	263	&	435	&	$p(t^{1})$	&	1.18	&	1.18	&	1.39	\\
	&	13 Aug. 2017	&	Io	&	434	&	435	&	$p(t^{1})$	&	1.04	&	1.15	&	1.19	\\
	&	19 Aug. 2017	&	Europa	&	287	&	439	&	$p(s)$	&	1.09	&	1.24	&	1.35	\\
	&	25 Aug. 2017	&	Europa	&	284	&	443	&	$p(s)$	&	1.35	&	1.3	&	1.75	\\
	&	20 Sep. 2017	&	Europa	&	254	&	460	&	$p(t^{1})+p(xy^{1})$	&	1.29	&	1.04	&	1.33	\\
	&	23 Sep. 2017	&	Io	&	264	&	462	&	$p(xy^{1})$	&	0.99	&	1.30	&	1.30	\\
	&	08 Oct. 2017	&	Europa	&	257	&	472	&	$p(xy^{1})$	&	1.3	&	1.3	&	1.69	\\
	&	20 Oct. 2017	&	Europa	&	227	&	480	&	$p(t^{1})$	&	1.06	&	1.2	&	1.28	\\
	&	30 Nov. 2017	&	Europa	&	260	&	507	&	$p(s)$	&	1.22	&	1.21	&	1.48	\\
	&	30 Nov. 2017	&	Io	&	267	&	507	&	$p(t^{1})+p(fwhm^{1})$	&	1.21	&	1.00	&	1.21	\\
	&	03 Dec. 2017	&	Io	&	262	&	509	&	$p(t^{1})$	&	1.13	&	1.37	&	1.55	\\
	&	03 Dec. 2017	&	Europa	&	259	&	509	&	$p(t^{1})$	&	1.04	&	1.00	&	1.04	\\
	&	06 Dec. 2017	&	Europa	&	212	&	511	&	$p(t^{1})$	&	1.89	&	1.00	&	1.89	\\
	&	28 Aug. 2017	&	Europa	&	154	&	445	&	$p(s)$	&	1.13	&	1.07	&	1.21	\\
	&	28 Aug. 2017	&	Io	&	156	&	445	&	$p(s)$	&	1.16	&	1.00	&	1.16	\\
\hline
c	&	28 Aug. 2017	&	Europa	&	178	&	294	&	$p(fwhm^{1})$	&	1.14	&	1.00	&	1.14	\\
	&	28 Aug. 2017	&	Io	&	272	&	294	&	$p(t^{1})$	&	1.10	&	1.61	&	1.76	\\
	&	14 Sep. 2017	&	Europa	&	247	&	301	&	$p(t^{1})$	&	1.08	&	1.35	&	1.45	\\
	&	15 Sep. 2017	&	Io	&	339	&	301	&	$p(t^{1})+p(a^{1})p(fwhm^{1})$	&	1.95	&	1.00	&	1.95	\\
	&	06 Oct. 2017	&	Europa	&	364	&	310	&	$p(t^{2})$	&	1.12	&	1.19	&	1.33	\\
	&	18 Oct. 2017	&	Europa	&	264	&	315	&	$p(t^{1})$	&	1.13	&	1.04	&	1.18	\\
	&	21 Nov. 2017	&	Europa	&	318	&	329	&	$p(b^{1})$	&	1.14	&	1.21	&	1.37	\\
	&	21 Nov. 2017	&	Io	&	265	&	329	&	$p(t^{1})+p(fwhm^{1})$	&	1.07	&	1.37	&	1.47	\\
	&	08 Dec. 2017	&	Europa	&	240	&	336	&	$p(s)$	&	1\textbf{.}11	&	1.18	&	1.31	\\
	&	08 Dec. 2017	&	Io	&	243	&	336	&	$p(a^{1})$	&	1.08	&	1.27	&	1.38	\\
	&	04 Nov. 2017	&	Europa	&	267	&	322	&	$p(t^{1})$	&	1.19	&	1.00	&	1.19	\\
\hline
    
d	&	26 Jul. 2017	&	Europa	&	422	&	72	&	$p(s)$	&	1.03	&	1.78	&	1.82	\\
	&	03 Aug. 2017	&	Europa	&	325	&	74	&	$p(t^{1})$	&	1.18	&	1.31	&	1.55	\\
	&	03 Aug. 2017	&	Io	&	378	&	74	&	$p(t^{1})+p(fwhm^{1})$	&	1.16	&	1.38	&	1.59	\\
	&	07 Aug. 2017	&	Europa	&	320	&	75	&	$p(t^{1})+p(fwhm^{1})$	&	1.17	&	1.00	&	1.17	\\
	&	07 Oct. 2017	&	Europa	&	322	&	90	&	$p(t^{1}) + p(xy^{1})$	&	1.07	&	1.13	&	1.21	\\ 

    \hline
e	&	29 Jun. 2017	&	Europa	&	422	&	45	&	$p(s)$	&	1.19	&	1.00	&	1.19	\\
	&	29 Jun. 2017	&	Io	&	401	&	45	&	$p(t^{1})$	&	1.06	&	1.33	&	1.41	\\
	&	05 Jul. 2017	&	Europa	&	448	&	46	&	$p(a^{1})+p(fwhm^{1})$	&	1.44	&	1.10	&	1.58	\\
	&	05 Jul. 2017	&	Io	&	445	&	46	&	$p(t^{2})+p(fwhm^{1})$	&	1.13	&	1.00	&	1.13	\\
	&	17 Aug. 2017	&	Europa	&	388	&	53	&	$p(s)$	&	0.93	&	1.39	&	1.30	\\
	&	17 Aug. 2017	&	Io	&	198	&	53	&	$p(s)$	&	0.91	&	1.05	&	0.95	\\
	&	23 Aug. 2017	&	Europa	&	418	&	54	&	$p(s)$	&	1.14	&	1.82	&	2.08	\\
	&	23 Aug. 2017	&	Io	&	415	&	54	&	$p(s)$	&	1.14	&	1.35	&	1.53	\\
\hline
f	&	27 Aug. 2017	&	Europa	&	363	&	35	&	$p(s)$	&	1.14	&	1.55	&	1.76	\\
	&	10 Oct. 2017	&	Europa	&	608	&	40	&	$p(s)$	&	1.11	&	1.42	&	1.58	\\
\hline
g	&	19 Jun. 2017	&	Europa	&	497	&	21	&	$p(fwhm^{1})$	&	0.95	&	1.05	&	1.00	\\
	&	26 Jul. 2017	&	Europa	&	475	&	22	&	$p(s)$	&	1.24	&	1.48	&	1.83	\\
	&	27 Jul. 2017	&	Europa	&	533	&	23	&	$p(s)$	&	1.23	&	1.08	&	1.34	\\
\hline
h	&	27 Jul. 2017	&	Europa	&	741	&	16	&	$p(a^{1})$	&	1.28	&	1.70	&	2.18	\\
	&	15 Aug. 2017	&	Io	&	412	&	17	&	$p(t^{1})$	&	1.01	&	1.08	&	1.19	\\
	&	15 Aug. 2017	&	Europa	&	434	&	17	&	$p(a^{1})$	&	0.97	&	1.81	&	1.77	\\
\end{longtable}
\begin{tablenotes}\footnotesize
\item \textbf{Notes.} For each light curve, this table shows the date of acquisition, the used instrument, the number of data points, the epoch based on the transit ephemeris presented in \citep{Delrez2018}, the selected baseline function (see Section.\ref{dataanalysis}) and the deduced values for $\beta_{w}$ , $\beta_{r}$ , and $CF= \beta_{r} * \beta_{w}$ (see Section.\ref{dataanalysis}). For the baseline function, p($\epsilon^{N}$ ) denotes, respectively, a N-order polynomial function of time ($\epsilon$ = t), the full width at half maximum ($\epsilon$ = fwhm), x and y positions ($\epsilon$ = xy), the background ($\epsilon$ = b), the airmass ($\epsilon$ = a) and a scalar ($\epsilon$ = s). 
\end{tablenotes}
\end{center}

\begin{center}
\begin{longtable}{|l|l|l|l|l|l|l|l|l|}
\caption{Same as Table \ref{baseline_spc}, but for K2.}
\label{baseline_k2} \\
\hline
\hline \multicolumn{1}{|c|}{\textbf{Planet}} & \multicolumn{1}{|c|}{\textbf{Date}} & \multicolumn{1}{|c|}{\textbf{Number of points}} & \multicolumn{1}{c|}{\textbf{Epoch}} & \multicolumn{1}{c|}{\textbf{Baseline}} & \multicolumn{1}{|c|}{\textbf{$\beta_{w}$}} & \multicolumn{1}{|c|}{\textbf{$\beta_{r}$}} & \multicolumn{1}{|c|}{\textbf{CF}} \\ \hline 
\endfirsthead

\multicolumn{3}{c}%
{{\bfseries \tablename\ \thetable{} -- continued from previous page}} \\
\\ \hline 
\endhead

\hline \multicolumn{3}{|r|}{{Continued on next page}} \\ \hline
\endfoot

\hline \hline
\endlastfoot
\hline
b	& 18 Dec. 2016	&	301	&	277	&	$p(t^{2})$	&	0.86	&	1.84	&	1.59	\\
    & 20 Dec. 2016	&	303	&	278	&	$p(t^{3})$	&	0.88	&	1.68	&	1.47	\\
    & 21 Dec. 2016	&	303	&	279	&	$p(t^{1})$	&	0.82	&	1.08	&	0.95	\\
    & 23 Dec. 2016	&	304	&	280	&	$p(t^{1})$	&	0.84	&	1.00	&	0.84	\\
    & 26 Dec. 2016	&	242	&	282	&	$p(s)$	&	0.91	&	1.11	&	1.01	\\
    & 27 Dec. 2016	&	241	&	283	&	$p(s)$	&	0.92	&	1.08	&	1.00	\\
    & 29 Dec. 2016	&	305	&	284	&	$p(t^{2})$	&	0.91	&	1.38	&	1.26	\\
    & 30 Dec. 2016	&	304	&	285	&	$p(s)$	&	0.84	&	1.34	&	1.13	\\
    & 01 Jan. 2017	&	303	&	286	&	$p(t^{2})$	&	0.86	&	1.01	&	0.87	\\
    & 02 Jan. 2017	&	305	&	287	&	$p(t^{1})$	&	0.90	&	1.74	&	1.57	\\
    & 04 Jan. 2017	&	303	&	288	&	$p(s)$	&	0.80	&	1.74	&	1.40	\\
    & 05 Jan. 2017	&	214	&	289	&	$p(t^{1})$	&	0.81	&	1.00	&	1.81	\\
    & 07 Jan. 2017	&	302	&	290	&	$p(t^{3})$	&	0.87	&	1.15	&	1.01	\\
    & 08 Jan. 2017	&	269	&	291	&	$p(t^{3})$	&	0.93	&	1.09	&	1.02	\\
    & 10 Jan. 2017	&	303	&	292	&	$p(s)$	&	0.87	&	1.82	&	1.57	\\
    & 11 Jan. 2017	&	303	&	293	&	$p(t^{3})$	&	0.84	&	1.07	&	0.91	\\
    & 13 Jan. 2017	&	305	&	294	&	$p(t^{1})$	&	0.89	&	1.12	&	1.00	\\
    & 14 Jan. 2017	&	305	&	295	&	$p(t^{2})$	&	0.90	&	1.28	&	1.16	\\
    & 16 Jan. 2017	&	297	&	296	&	$p(s)$	&	0.91	&	1.63	&	1.49	\\
    & 17 Jan. 2017	&	215	&	297	&	$p(t^{1})$	&	0.84	&	1.53	&	1.28	\\
    & 19 Jan. 2017	&	206	&	298	&	$p(s)$	&	0.82	&	1.68	&	1.39	\\
    & 20 Jan. 2017	&	259	&	299	&	$p(s)$	&	0.92	&	1.22	&	1.13	\\
    & 22 Jan. 2017	&	304	&	300	&	$p(t^{1})$	&	0.88	&	1.48	&	1.32	\\
    & 23 Jan. 2017	&	303	&	301	&	$p(t^{4})$	&	0.89	&	1.00	&	0.89	\\
    & 25 Jan. 2017	&	302	&	302	&	$p(s)$	&	0.82	&	1.19	&	0.87	\\
    & 26 Jan. 2017	&	302	&	303	&	$p(t^{1})$	&	0.86	&	1.43	&	1.23 \\
    & 29 Jan. 2017	&	293	&	305	&	$p(t^{2})$	&	0.87	&	1.04	&	0.91 \\
    & 31 Jan. 2017	&	304	&	306	&	$p(t^{3})$	&	0.90	&	1.22	&	1.11 \\
    & 07 Feb. 2017	&	306	&	311	&	$p(t^{3})$	&	0.81	&	1.09	&	0.87 \\
    & 10 Feb. 2017	&	300	&	313	&	$p(s)$	&	0.97	&	1.63	&	1.58 \\
    & 12 Feb. 2017	&	304	&	314	&	$p(s)$	&	1.04	&	1.31	&	1.36 \\
    & 13 Feb. 2017	&	302	&	315	&	$p(t^{4})$	&	0.92	&	1.12	&	1.03 \\
    & 15 Feb. 2017	&	304	&	316	&	$p(t^{2})$	&	0.94	&	1.34	&
    1.26 \\
    & 16 Feb. 2017	&	303	&	317	&	$p(t^{3})$	&	0.94	&	1.16	&	1.09 \\
    & 18 Feb. 2017	&	296	&	318	&	$p(t^{1})$	&	0.81	&	1.09	&	0.87 \\
    & 19 Feb. 2017	&	305	&	319	&	$p(t^{1})$	&	0.88	&	1.11	&	0.98 \\
    & 21 Feb. 2017	&	206	&	320	&	$p(s)$	&	0.91	&	1.54	&	1.40 \\
    & 24 Feb. 2017	&	294	&	322	&	$p(t^{1})$	&	0.95	&	1.08	&	1.02 \\
    & 26 Feb. 2017	&	305	&	323	&	$p(t^{3})$	&	0.87	&	1.00	&	0.87 \\
    & 01 Mar. 2017	&	196	&	325	&	$p(s)$	&	0.95	&	1.19	&	1.13 \\
    & 01 Mar. 2017	&	291	&	326	&	$p(t^{1})$	&	0.93	&	1.00	&	0.93 \\
    & 04 Mar. 2017	&	305	&	327	&	$p(s)$	&	1.02	&	1.89	&	1.93 \\
    \hline
c   & 18 Dec. 2016	&	304	&	189	&	$p(t^{1})$	&	0.83	&	1.00	&	0.83 \\
& 20 Dec. 2016	&	219	&	190	&	$p(t^{2})$	&	0.87	&	1.28	&	
1.07 \\
& 22 Dec. 2016	&	217	&	191	&	$p(s)$	&	0.81	&	1.73	&	1.41 \\ 
& 25 Dec. 2016	&	304	&	192	&	$p(s)$	&	0.86	&	1.64	&	1.41 \\
& 27 Dec. 2016	&	238	&	193	&	$p(s)$	&	0.83	&	1.00	&	0.83 \\
& 30 Dec. 2016	&	303	&	194	&	$p(t^{1})$	&	0.80	&	1.30	&	1.04 \\
& 03 Jan. 2017	&	232	&	196	&	$p(s)$	&	0.89	&	2.14	&	1.90 
\\
& 05 Jan. 2017	&	185	&	197	&	$p(t^{1})$	&	0.89	&	1.06	&	0.94 
\\
& 07 Jan. 2017	&	250	&	198	&	$p(t^{4})$	&	0.88	&	1.22	&	1.08 
\\
& 11 Jan. 2017	&	304	&	199	&	$p(s)$	&	0.85	&	1.51	&	1.28 
\\
& 13 Jan. 2017	&	302	&	200	&	$p(s)$	&	0.84	&	1.35	&	1.14 
\\
& 16 Jan. 2017	&	249	&	201	&	$p(t^{2})$	&	0.81	&	1.25	&	1.03
\\
& 18 Jan. 2017	&	244	&	202	&	$p(s)$	&	0.80	&	1.09	&	0.87
\\
& 20 Jan. 2017	&	284	&	203	&	$p(t^{1})$	&	0.84	&	1.17	&	0.98
\\
& 23 Jan. 2017	&	305	&	204	&	$p(t^{3})$	&	0.86	&	1.00	&	0.86
\\
& 25 Jan. 2017	&	304	&	205	&	$p(s)$	&	0.91	&	1.46	&	1.34
\\
& 27 Jan. 2017	&	233	&	206	&	$p(s)$	&	0.84	&	1.29	&	1.08
\\
& 30 Jan. 2017	&	216	&	207	&	$p(t^{1})$	&	0.91	&	1.13	&	1.03
\\
& 06 Feb. 2017	&	188	&	210	&	$p(t^{3})$	&	0.85	&	1.00	&	0.85
\\
& 09 Feb. 2017	&	221	&	211	&	$p(t^{1})$	&	0.87	&	1.31	&	1.14
\\
& 11 Feb. 2017	&	303	&	212	&	$p(t^{2})$	&	0.88	&	1.18	&	1.05
\\
& 14 Feb. 2017	&	304	&	213	&	$p(t^{3})$	&	0.85	&	1.77	&	1.51
\\
& 16 Feb. 2017	&	258	&	214	&	$p(t^{2})$	&	0.95	&	1.69	&	1.60
\\
& 18 Feb. 2017	&	253	&	215	&	$p(t^{3})$	&	0.85	&	1.11	&	1.94
\\
& 21 Feb. 2017	&	210	&	216	&	$p(t^{1})$	&	0.92	&	1.42	&	
1.31
\\
& 23 Feb. 2017	&	307	&	217	&	$p(t^{2})$	&	0.89	&	1.31	&	1.17
\\  
& 26 Feb. 2017	&	304	&	218	&	$p(s)$	&	0.89	&	2.00	&	1.79
\\ 
& 28 Feb. 2017	&	306	&	219	&	$p(t^{2})$	&	0.93	&	1.00	&	0.93
\\  
& 03 Mar. 2017	&	305	&	220	&	$p(t^{3})$	&	0.87	&	1.00	&	0.87
\\  
\hline
d & 16 Dec. 2016	&	305	&	44	&	$p(s)$	&	0.84	&	1.13	&	0.96
\\ 
 & 20 Dec. 2016	&	203	&	45	&	$p(t^{4})$	&	0.79	&	1.00	&	0.79
\\
 & 28 Dec. 2016	&	304	&	47	&	$p(t^{4})$	&	0.88	&	1.13	&	1.00
\\
& 01 Jan. 2017	&	186	&	48	&	$p(t^{1})$	&	0.83	&	1.00	&	0.83
\\
& 05 Jan. 2017	&	198	&	49	&	$p(s)$	&	0.89	&	1.01	&	0.90
\\
& 09 Jan. 2017	&	305	&	50	&	$p(t^{3})$	&	0.79	&	1.00	&	0.79
\\
& 13 Jan. 2017	&	304	&	51	&	$p(t^{1})$	&	0.84	&	1.09	&	0.91
\\
 &17 Jan. 2017	&	491	&	52	&	$p(s)$	&	0.91	&	1.48	&	1.35
\\
& 21 Jan. 2017	&	306	&	53	&	$p(t^{1})$	&	0.87	&	1.30	&	1.13
\\
& 25 Jan. 2017	&	298	&	54	&	$p(t^{3})$	&	0.87	&	1.45	&	1.27
\\
& 07 Feb. 2017	&	210	&	57	&	$p(s)$	&	0.87	&	1.00	&	0.87
\\
& 23 Feb. 2017	&	305	&	61	&	$p(t^{1})$	&	0.87	&	1.11	&	0.97
\\
& 27 Feb. 2017	&	304	&	61	&	$p(t^{1})$	&	0.93	&	1.40	&	1.30
\\
& 03 Mar. 2017	&	306	&	63	&	$p(s)$	&	0.97	&	1.00	&	0.97
\\
\hline
e   & 17 Dec. 2016	&	259	&	70	&	$p(t^{1})$	&	0.84	&	1.40	&	1.17 \\
& 23 Dec. 2016	&	303	&	71	&	$p(t^{1})$	&	0.87	&	1.27	&	
1.11 \\
& 04 Jan. 2016	&	296	&	73	&	$p(t^{1})$	&	0.88	&	2.01	&	1.78 \\
& 10 Jan. 2016	&	251	&	74	&	$p(t^{1})$	&	0.89	&	1.20	&	1.08 \\
& 16 Jan. 2016	&	306	&	75	&	$p(t^{1})$	&	0.87	&	1.04	&	0.90 \\
& 22 Jan. 2016	&	304	&	76	&	$p(t^{2})$	&	0.83	&	1.05	&	0.89 \\
& 28 Jan. 2016	&	304	&	77	&	$p(t^{1})$	&	0.91	&	1.00	&	
0.91 \\
& 10 Feb. 2016	&	304	&	79	&	$p(t^{1})$	&	0.90	&	1.55	&	1.40 \\
\hline
f   & 22 Dec. 2016	&	260	&	8	&	$p(s)$	&	0.90	&	1.52	&	1.37 \\
& 31 Dec. 2016	&	304	&	9	&	$p(s)$	&	0.88	&	1.16	&	1.03 \\
& 09 Jan. 2017	&	304	&	10	&	$p(s)$	&	0.90	&	1.79	&	1.62 \\
& 19 Jan. 2017	&	223	&	11	&	$p(t^{1})$	&	0.89	&	1.15	&	
1.03 \\
& 15 Feb. 2017	&	303	&	14	&	$p(s)$	&	0.87	&	1.67	&	1.46 \\
& 15 Feb. 2017	&	301	&	15	&	$p(s)$	&	0.89	&	1.30	&	1.15 \\
\hline
g   & 10 Jan. 2017	&	199	&	8	&	$p(s)$	&	0.89	&	1.05	&	0.93 \\
& 16 Feb. 2017	&	256	&	11	&	$p(t^{1})$	&	0.88	&	1.51	&	1.34 \\
& 01 Mar. 2017	&	156	&	12	&	$p(s)$	&	0.96	&	1.41	&	1.35 \\
\hline
h   & 02 Jan. 2017	&	304 &	5	&	$p(t^{1})$	&	0.82	&	1.06	&	0.88 \\
\hline

\end{longtable}
\end{center}

\begin{center}
\begin{longtable}{|l|l|l|l|l|l|l|l|l|}
\caption{Same as Table \ref{baseline_spc}, but for LT.} 
\label{baseline_liverpool} \\
\hline
\hline \multicolumn{1}{|c|}{\textbf{Planet}} & \multicolumn{1}{|c|}{\textbf{Date}}  & \multicolumn{1}{|c|}{\textbf{Number of points}} & \multicolumn{1}{c|}{\textbf{Epoch}} & \multicolumn{1}{c|}{\textbf{Baseline}} & \multicolumn{1}{|c|}{\textbf{$\beta_{w}$}} & \multicolumn{1}{|c|}{\textbf{$\beta_{r}$}} & \multicolumn{1}{|c|}{\textbf{CF}} \\ \hline 
\endfirsthead

\multicolumn{3}{c}%
{{\bfseries \tablename\ \thetable{} -- continued from previous page}} \\
\\ \hline 
\endhead

\hline \multicolumn{3}{|r|}{{Continued on next page}} \\ \hline
\endfoot

\hline \hline
\endlastfoot
\hline
b	& 31 May. 2017	&	139	&	386	&	$p(t^{1})+p(fwhm^{1})$	&	1.23	&	1.00	&	1.23	\\
	& 23 Jul. 2017	&	152	&	421	&	$p(s)$	&	1.00
	&	1.09	&	1.09	\\
    & 29 Jul. 2017	&	153	&	425	&	$p(s)$	&	0.99
	&	1.08	&	1.07	\\
    & 5 Aug. 2017	&	156	&	429	&	$p(t^{1})$	&	1.58
	&	1.00	&	1.58	\\
\hline
c	& 01 Jul. 2017	&	157	&	270	&	$p(s)$	&   0.88	&	1.43	&	1.26	\\
	& 07 Sep. 2017	&	178	&	298	&	$p(t^{1}))$	&	0.95
    &	1.00	&	0.95	\\
    & 19 Sep. 2017	&	178	&	303	&	$p(t^{1})$	&	1.31
    &	1.24	&	1.63	\\
    & 28 Oct. 2017	&	176	&	319	&	$p(s)$	&	1.11    &	1.31	&	1.46	\\
    & 5 Aug. 2017	&	187	&	284	&	$p(s)$	&	1.51
	&	1.25	&	1.79	\\
\hline
d	& 21 Sep. 2017	&	227	&	113	&	$p(t^{1})+p(fwhm^{1})$	&  1.45
&	1.05	&	1.52	\\
\hline
e	& 17 Aug. 2017	&	274	&	110	&	$p(t^{1})$	&  1.30
&	1.28	&	1.66	\\
	& 17 Aug. 2017	&	202	&	118	&	$p(s)$	&  1.00
&	1.55	&	1.55	\\
\hline
h	& 15 Aug. 2017	&	378	&	17	&	$p(t^{1})$	&  1.00
&	1.00	&	1.00	\\

\end{longtable}
\end{center}

\newpage
\label{AppendixB}
\section{Results from the individual analysis}

\begin{center}
\begin{longtable}{|l|l|ll|ll|}
\caption{Transit timings and depths obtained from the individual analyses of SPECULOOS light curves. Each row represents a transit, the first column gives the planet's name, the second the epoch of the transit, the third the mid-transit timing and the corresponding error resulting from the analysis and the last column the transit depth and corresponding error resulting from the analysis.}  \label{individual_spc} \\
\hline
\hline \multicolumn{1}{|c|}{\textbf{Planet}} & \multicolumn{1}{|c|}{\textbf{Epoch}} & \multicolumn{2}{c|}{\textbf{Transit timing [$BJD_{TDB}-2450000$]}} & \multicolumn{2}{c|}{\textbf{Transit depth (\%)}} \\ \hline 
\endfirsthead

\multicolumn{3}{c}%
{{\bfseries \tablename\ \thetable{} -- continued from previous page}} \\
\\ \hline 
\endhead

\hline \multicolumn{3}{|r|}{{Continued on next page}} \\ \hline
\endfoot

\hline \hline
\endlastfoot
\hline
b	&	398	&	7923.84586	&	0.00043	&	0.764	&	0.060	\\
	&	406	&	7935.93284	&	0.00028	&	0.842	&	0.047	\\
	&	406	&	7935.93316	&	0.00053	&	0.893	&	0.088	\\
	&	427	&	7967.66254	&	0.00053	&	0.686	&	0.068	\\
	&	431	&	7973.70588	&	0.00058	&	0.759	&	0.078	\\
	&	435	&	7979.74899	&	0.00030	&	0.835	&	0.058	\\
	&	435	&	7979.74864	&	0.00034	&	0.738	&	0.048	\\
	&	439	&	7985.79209	&	0.00034	&	0.721	&	0.052	\\
	&	443	&	7991.83579	&	0.00041	&	0.845	&	0.079	\\
	&	460	&	8017.52106	&	0.00041	&	0.774	&	0.079	\\
	&	462	&	8020.54219	&	0.00036	&	0.758	&	0.056	\\
	&	472	&	8035.65192	&	0.00065	&	0.801	&	0.085	\\
	&	480	&	8047.73788	&	0.00059	&	0.676	&	0.094	\\
	&	507	&	8088.53228	&	0.00033	&	0.796	&	0.060	\\
	&	507	&	8088.53206	&	0.00026	&	0.920	&	0.059	\\
	&	509	&	8091.55411	&	0.00036	&	0.878	&	0.065	\\
	&	509	&	8091.55364	&	0.00035	&	0.809	&	0.045	\\
	&	511	&	8094.57595	&	0.00067	&	0.822	&	0.120	\\
	&	445	&	7994.85842	&	0.00047	&	0.819	&	0.084	\\
	&	445	&	7994.85833	&	0.00051	&	0.855	&	0.083	\\
 \hline
c	&	294	&	7994.81758	&	0.0004	&	0.835	&	0.068	\\
	&	294	&	7994.81885	&	0.00065	&	0.695	&	0.082	\\
	&	301	&	8011.77150	&	0.00046	&	0.826	&	0.066	\\
	&	301	&	8011.77102	&	0.00036	&	0.878	&	0.078	\\
	&	310	&	8033.56743	&	0.00041	&	0.801	&	0.060	\\
	&	315	&	8045.67598	&	0.00035	&	0.738	&	0.055	\\
	&	329	&	8079.58077	&	0.00042	&	0.649	&	0.055	\\
	&	329	&	8079.58172	&	0.00050	&	0.679	&	0.055	\\
	&	336	&	8096.53342	&	0.00037	&	0.789	&	0.055	\\
	&	336	&	8096.53330	&	0.00051	&	0.819	&	0.062	\\
	&	322	&	8062.62794	&	0.00039	&	0.727	&	0.160	\\
\hline
d	&	72	&	7961.73755	&	0.00012	&	0.394	&	0.057	\\
	&	74	&	7969.83771	&	0.00020	&	0.264	&	0.062	\\
	&	74	&	7969.83665	&	0.00100	&	0.375	&	0.065	\\
	&	75	&	7973.88834	&	0.00140	&	0.401	&	0.062	\\
	&	90	&	8034.62829	&	0.00063	&	0.405	&	0.048	\\
 \hline
e	&	45	&	7934.83251	&	0.00088	&	0.442	&	0.046	\\
	&	45	&	7934.82990	&	0.00092	&	0.417	&	0.044	\\
	&	46	&	7940.93132	&	0.00049	&	0.547	&	0.048	\\
	&	46	&	7940.92923	&	0.00061	&	0.454	&	0.055	\\
	&	53	&	7983.62886	&	0.00095	&	0.522	&	0.055	\\
	&	53	&	7983.62706	&	0.00053	&	0.590	&	0.057	\\
	&	54	&	7989.73173	&	0.00210	&	0.449	&	0.065	\\
	&	54	&	7989.72916	&	0.00067	&	0.458	&	0.045	\\
 \hline
f   &   35  & 	7993.63410  &	0.00070	&	0.741	&	0.074	\\
	&	40	&	8039.66021	&	0.00084 &	0.639	&	0.056	\\
 \hline
g	&	21	&	7924.76924	&	0.00055	&	0.791	&	0.051	\\
	&	24	&	7961.82599	&	0.00075	&	0.723	&	0.059	\\
	&	29	&	7813.60697	&	0.00200	&	0.867	&	0.17	\\
 \hline
h	&	16	&	7962.86330	&	0.0018 &	0.372	&	0.052	\\
	&	17	&	7981.63159	&	0.0016	&	0.290	&	0.046	\\
	&	17	&	7981.63059	&	0.0030	&	0.301	&	0.046	\\
 
\end{longtable}
\end{center}

\begin{center}
\begin{longtable}{|l|l|ll|ll|}
\caption{Transit timings and depths obtained from the individual analyses of K2 light curves. Each row represents a transit, the first column gives the planet's name, the second the epoch of the transit, the third the mid-transit timing and the corresponding error resulting from the analysis and the last column the transit depth and corresponding error resulting from the analysis.} \\
\hline
\hline \multicolumn{1}{|c|}{\textbf{Planet}} & \multicolumn{1}{|c|}{\textbf{Epoch}} & \multicolumn{2}{c|}{\textbf{Transit timing [$BJD_{TDB}-2450000$]}} & \multicolumn{2}{c|}{\textbf{Transit depth (\%)}} \\ \hline 
\endfirsthead

\multicolumn{3}{c}%
{{\bfseries \tablename\ \thetable{} -- continued from previous page}} \\
\endhead

\hline \multicolumn{3}{|r|}{{Continued on next page}} \\ \hline
\endfoot

\hline \hline
\endlastfoot
\hline
b & 277 & 7741.02841 & 0.0011 & 0.959 & 0.200  \\
 & 278 & 7742.54031 & 0.00120 & 0.804 & 0.160  \\
 & 279 & 7744.05191 & 0.00063 & 0.740 & 0.095  \\
 & 280 & 7745.56254 & 0.00071 & 0.721 & 0.080  \\
 & 282 & 7748.58511 & 0.00071 & 0.728 & 0.084  \\
 & 283 & 7750.09533 & 0.00150 & 0.776 & 0.110  \\
 & 284 & 7751.60539 & 0.00093 & 0.799 & 0.150  \\
 & 285 & 7753.11716 & 0.00064 & 0.746 & 0.100  \\
 & 286 & 7754.62846 & 0.00071 & 0.720 & 0.089  \\
 & 287 & 7756.13952 & 0.00110 & 0.775 & 0.150  \\
 & 288 & 7757.64925 & 0.00098 & 0.784 & 0.100  \\
 & 289 & 7759.16120 & 0.00100 & 0.689 & 0.080  \\
 & 290 & 7760.67229 & 0.00086 & 0.743 & 0.097  \\
 & 291 & 7762.18295 & 0.00090 & 0.569 & 0.055  \\
 & 292 & 7763.69272 & 0.00110 & 0.741 & 0.130  \\
 & 293 & 7765.20352 & 0.00056 & 0.843 & 0.083  \\
 & 294 & 7766.71525 & 0.00074 & 0.766 & 0.089  \\
 & 295 & 7768.22451 & 0.00089 & 0.932 & 0.180  \\
 & 296 & 7769.73779 & 0.00140 & 0.666 & 0.200  \\
 & 297 & 7771.24857 & 0.00140 & 0.673 & 0.150  \\
 & 298 & 7772.75851 & 0.00120 & 0.643 & 0.120  \\
 & 299 & 7774.26913 & 0.00085 & 0.889 & 0.110  \\
 & 300 & 7775.78022 & 0.00099 & 0.736 & 0.120  \\
 & 301 & 7777.28984 & 0.00069 & 0.685 & 0.085  \\
 & 302 & 7778.80191 & 0.00084 & 0.632 & 0.070  \\
 & 303 & 7780.31394 & 0.00058 & 0.719 & 0.089  \\
 & 305 & 7783.33438 & 0.00110 & 0.604 & 0.082  \\
 & 306 & 7784.84448 & 0.00150 & 0.555 & 0.110  \\
 & 311 & 7792.40048 & 0.00110 & 0.788 & 0.092  \\
 & 313 & 7795.42062 & 0.00110 & 0.902 & 0.210  \\
 & 314 & 7796.93214 & 0.00093 & 0.772 & 0.130  \\
 & 315 & 7798.44260 & 0.00065 & 0.836 & 0.120  \\
 & 316 & 7799.95368 & 0.00100 & 0.822 & 0.200  \\
 & 317 & 7801.46362 & 0.00099 & 0.707 & 0.100  \\
 & 318 & 7802.97696 & 0.00099 & 0.830 & 0.280  \\
 & 319 & 7804.48723 & 0.00065 & 0.783 & 0.099  \\
 & 320 & 7805.99725 & 0.00110 & 0.669 & 0.160  \\
 & 322 & 7809.02001 & 0.00063 & 0.988 & 0.120  \\
 & 323 & 7810.52858 & 0.00059 & 0.809 & 0.120  \\
 & 325 & 7813.55299 & 0.00079 & 0.866 & 0.130  \\
 & 326 & 7815.06305 & 0.00067 & 0.693 & 0.073  \\
 & 327 & 7816.57407 & 0.00058 & 0.851 & 0.086  \\
 \hline
c & 189 & 7740.53417 & 0.00083 & 0.589 & 0.091  \\
 & 190 & 7742.95370 & 0.00100 & 0.737 & 0.091  \\
 & 191 & 7745.37836 & 0.00200 & 0.656 & 0.150  \\
 & 192 & 7747.79745 & 0.00100 & 0.864 & 0.150  \\
 & 193 & 7750.21906 & 0.00092 & 0.699 & 0.065  \\
 & 194 & 7752.64173 & 0.00100 & 0.652 & 0.079  \\
 & 196 & 7757.48363 & 0.00150 & 0.770 & 0.160  \\
 & 197 & 7759.90355 & 0.00081 & 0.552 & 0.077  \\
 & 198 & 7762.32917 & 0.00098 & 0.697 & 0.100  \\
 & 199 & 7764.74926 & 0.00120 & 0.818 & 0.120  \\
 & 200 & 7767.17041 & 0.00120 & 0.791 & 0.160  \\
 & 201 & 7769.59305 & 0.00082 & 0.579 & 0.090  \\
 & 202 & 7772.01577 & 0.00110 & 0.846 & 0.081  \\
 & 203 & 7774.43531 & 0.00084 & 0.732 & 0.090  \\
 & 204 & 7776.85884 & 0.00084 & 0.789 & 0.130  \\
 & 205 & 7779.27985 & 0.00150 & 0.713 & 0.110  \\
 & 206 & 7781.70135 & 0.00081 & 0.785 & 0.081  \\
 & 207 & 7784.12337 & 0.00080 & 0.837 & 0.100  \\
 & 210 & 7791.38904 & 0.00080 & 0.588 & 0.086  \\
 & 211 & 7793.81167 & 0.00085 & 0.674 & 0.082  \\
 & 212 & 7796.23257 & 0.00072 & 0.771 & 0.085  \\
 & 213 & 7798.65449 & 0.00110 & 0.798 & 0.140  \\
 & 214 & 7801.07700 & 0.00084 & 0.771 & 0.140  \\
 & 215 & 7803.49803 & 0.00100 & 0.604 & 0.090  \\
 & 216 & 7805.91971 & 0.00068 & 0.686 & 0.080  \\
 & 217 & 7808.34120 & 0.00120 & 0.797 & 0.120  \\
 & 218 & 7810.76238 & 0.00210 & 0.809 & 0.400  \\
 & 219 & 7813.18452 & 0.00110 & 0.663 & 0.071  \\
 & 220 & 7815.60631 & 0.00070 & 0.856 & 0.074  \\
\hline
d & 17 & 7738.99254 & 0.00400 & 0.286 & 0.110 \\
 & 18 & 7743.03818 & 0.00120 & 0.564 & 0.092 \\
 & 20 & 7751.14013 & 0.00180 & 0.468 & 0.100 \\
 & 21 & 7755.18855 & 0.00140 & 0.537 & 0.120 \\
 & 22 & 7759.24739 & 0.00180 & 0.461 & 0.073 \\
 & 23 & 7763.28944 & 0.00130 & 0.419 & 0.062 \\
 & 24 & 7767.34079 & 0.00330 & 0.318 & 0.130 \\
 & 25 & 7771.39074 & 0.00420 & 0.453 & 0.120 \\
 & 26 & 7775.44035 & 0.00180 & 0.466 & 0.090 \\
 & 27 & 7779.48982 & 0.00320 & 0.603 & 0.240 \\
 & 30 & 7791.64154 & 0.00098 & 0.570 & 0.076 \\
 & 34 & 7807.84073 & 0.00570 & 0.304 & 0.130 \\
 & 35 & 7811.88917 & 0.00460 & 0.412 & 0.210 \\
 & 36 & 7815.94153 & 0.00170 & 0.361 & 0.110 \\
 \hline
e & 13 & 7739.67183 & 0.00160 & 0.509 & 0.100 \\
 & 14 & 7745.77293 & 0.00180 & 0.514 & 0.110 \\
 & 16 & 7757.96796 & 0.00310 & 0.587 & 0.110 \\
 & 17 & 7764.07021 & 0.00150 & 0.521 & 0.120 \\
 & 18 & 7770.17149 & 0.00240 & 0.447 & 0.130 \\
 & 19 & 7776.26457 & 0.00190 & 0.383 & 0.075 \\
 & 20 & 7782.36274 & 0.00190 & 0.430 & 0.070 \\
 & 22 & 7794.56245 & 0.00180 & 0.599 & 0.089 \\
 \hline
f & 8 & 7745.03067 & 0.00210 & 0.613 & 0.160 \\
 & 9  & 7754.23474 & 0.00140 & 0.653 & 0.110 \\
 & 10 & 7763.44545 & 0.00240 & 0.651 & 0.130 \\
 & 11 & 7772.64854 & 0.00180 & 0.461 & 0.061 \\
 & 14 & 7800.27394 & 0.00220 & 0.524 & 0.120 \\
 & 15 & 7809.47737 & 0.00270 & 0.494 & 0.090 \\
 \hline
g & 8 & 7764.19229 & 0.00180 & 0.559 & 0.071  \\
 & 11 & 7801.25085 & 0.00120 & 0.727 & 0.100  \\
 & 12 & 7813.60698 & 0.00200 & 0.867 & 0.170  \\
 \hline
h & 5 & 7756.38806 & 0.00300 & 0.346 & 0.058 \\
 \label{individual_k2}
\end{longtable}
\end{center}

\begin{center}
\begin{longtable}{|l|l|ll|ll|}
\caption{Transit timings and depths obtained from the individual analyses of LT light curves. Each row represents a transit, the first column gives the planet's name, the second the epoch of the transit, the third the mid-transit timing and the corresponding error resulting from the analysis and the last column the transit depth and corresponding error resulting from the analysis.}  \label{individual_lt} \\
\hline
\hline \multicolumn{1}{|c|}{\textbf{Planet}} & \multicolumn{1}{|c|}{\textbf{Epoch}} & \multicolumn{2}{c|}{\textbf{Transit timing [$BJD_{TDB}-2450000$]}} & \multicolumn{2}{c|}{\textbf{Transit depth (\%)}} \\ \hline 
\endfirsthead

\multicolumn{3}{c}%
{{\bfseries \tablename\ \thetable{} -- continued from previous page}} \\
\endhead

\hline \multicolumn{3}{|r|}{{Continued on next page}} \\ \hline
\endfoot

\hline \hline
\endlastfoot
\hline
b & 386 & 7905.71514  & 0.00088 & 0.848 & 0.130  \\
& 421 & 7958.59599  & 0.00038& 0.696 & 0.062  \\
& 425 & 7964.63878 & 0.00043 & 0.830 & 0.063  \\
& 429 & 7970.68530  & 0.00051 & 0.706 & 0.063  \\
\hline
  c & 270 & 7936.69651  & 0.00040 & 0.721 & 0.053  \\
  & 298 & 8004.50488  & 0.00052 & 0.879 & 0.058  \\
  & 303 & 8016.61384  & 0.00087 & 0.612 & 0.090  \\
  & 319 & 8055.36295  & 0.00044 & 0.765 & 0.059   \\
  & 284 & 7970.60046  & 0.00085 & 0.638 & 0.070   \\
\hline
d & 86 & 8018.43071  & 0.00096 & 0.353 & 0.027  \\
\hline
  e & 53 & 7983.62882  & 0.00140 & 0.481 & 0.075  \\
  & 56 & 8032.43398  & 0.00180 & 0.475 & 0.100  \\
\hline
h & 17 & 7981.63343   & 0.00110 & 0.257  & 0.035  \\
 \label{appendix:indiv_liverpool}
\end{longtable}
\end{center}

\newpage

\label{AppendixC}

\section{Results from the global analysis}

\begin{center}
\begin{longtable}{|l|l|ll|ll|}
\caption{Median values and 1-$\sigma$ limits of the posterior PDFs deduced for the timings and depths from their global analyses for SPECULOOS observations. Each row represents a transit, the first column gives the planet's name, the second the epoch of the transit, the third the mid-transit timing and the corresponding error resulting from the analysis and the last column the transit depth and corresponding error resulting from the analysis.} \label{global_spc} \\
\hline
\hline \multicolumn{1}{|c|}{\textbf{Planet}} & \multicolumn{1}{|c|}{\textbf{Epoch}} & \multicolumn{2}{|c|}{\textbf{Transit timing [$BJD_{TDB}-2450000$]}} & \multicolumn{2}{c|}{\textbf{Transit depth (\%)}} \\ \hline 
\endfirsthead

\multicolumn{3}{c}%
{{\bfseries \tablename\ \thetable{} -- continued from previous page}} \\
\\ \hline 
\endhead

\hline \multicolumn{3}{|r|}{{Continued on next page}} \\ \hline
\endfoot

\hline \hline
\endlastfoot

\hline
b	&	398	&	7923.84588	&	0.00043	&	0.744	&	0.053	\\
	&	406	&	7935.93286	&	0.00023	&	0.882	&	0.040	\\
	&	406	&	7935.93286	&	0.00023	&	0.904	&	0.084	\\
	&	427	&	7967.66246	&	0.00054	&	0.706	&	0.090	\\
	&	431	&	7973.70578	&	0.00053	&	0.756	&	0.066	\\
	&	435	&	7979.74887	&	0.00022	&	0.852	&	0.052	\\
	&	435	&	7979.74887	&	0.00022	&	0.763	&	0.044	\\
	&	439	&	7985.79210	&	0.00031	&	0.737	&	0.047	\\
	&	443	&	7991.83581	&	0.00042	&	0.864	&	0.073	\\
	&	460	&	8017.52101	&	0.00061	&	0.758	&	0.072	\\
	&	472	&	8035.65154	&	0.00062	&	0.773	&	0.073	\\
	&	480	&	8047.73785	&	0.00061	&	0.788	&	0.065	\\
	&	462	&	8020.54220	&	0.0004	&	0.698	&	0.120	\\
	&	507	&	8088.53214	&	0.00022	&	0.809	&	0.051	\\
	&	507	&	8088.53214	&	0.00022	&	0.932	&	0.054	\\
	&	509	&	8091.55387	&	0.00026	&	0.895	&	0.059	\\
	&	509	&	8091.55387	&	0.00026	&	0.848	&	0.041	\\
	&	511	&	8094.57599	&	0.00059	&	0.82	&	0.110	\\
	&	445	&	7994.85799	&	0.00055	&	0.735	&	0.073	\\
	&	445	&	7994.85799	&	0.00055	&	0.784	&	0.078	\\
 \hline
c	&	294	&	7994.81840	&	0.00034	&	0.792	&	0.069	\\
	&	294	&	7994.81840	&	0.00034	&	0.684	&	0.078	\\
	&	301	&	8011.77116	&	0.00029	&	0.800	&	0.072	\\
	&	301	&	8011.77116	&	0.00029	&	0.904	&	0.076	\\
	&	310	&	8033.56743	&	0.00038	&	0.816	&	0.061	\\
	&	315	&	8045.67601	&	0.00034	&	0.73	&	0.050	\\
	&	329	&	8079.58130	&	0.00030	&	0.634	&	0.046	\\
	&	329	&	8079.58130	&	0.00030	&	0.67	&	0.044	\\
	&	336	&	8096.53332	&	0.00030	&	0.813	&	0.046	\\
	&	336	&	8096.53332	&	0.00030	&	0.818	&	0.056	\\
	&	322	&	8062.62799	&	0.00037	&	0.727	&	0.051	\\
 \hline
d	&	72	&	7961.73774	&	0.00130	&	0.398	&	0.061	\\
	&	74	&	7969.83692	&	0.00070	&	0.266	&	0.044	\\
	&	74	&	7969.83692	&	0.00070	&	0.376	&	0.053	\\
	&	75	&	7973.88758	&	0.00150	&	0.372	&	0.059	\\
	&	90	&	8034.62829	&	0.00069	&	0.409	&	0.050	\\
 \hline
e	&	45	&	7934.83078	&	0.00065	&	0.406	&	0.048	\\
	&	45	&	7934.83078	&	0.00065	&	0.421	&	0.038	\\
	&	46	&	7940.92999	&	0.00069	&	0.540	&	0.050	\\
	&	46	&	7940.92999	&	0.00069	&	0.471	&	0.057	\\
	&	53	&	7983.62772	&	0.00086	&	0.518	&	0.047	\\
	&	53	&	7983.62772	&	0.00086	&	0.553	&	0.070	\\
	&	54	&	7989.72944	&	0.00075	&	0.446	&	0.061	\\
	&	54	&	7989.72944	&	0.00075	&	0.463	&	0.049	\\
 \hline
f	&	35	&	7993.63412	&	0.00084	&	0.732	&	0.071	\\
	&	40	&	8039.66014	&	0.00091 &	0.653	&	0.055	\\
\hline
g	&	21	&	7924.76918	&	0.00140	&	0.810	&	0.092	\\
	&	24	&	7961.82610	&	0.00053	&	0.723	&	0.036	\\
	&	29	&	8060.65579	&	0.00047	&	0.758	&	0.036	\\
 \hline
h	&	16	&	7962.86307	&	0.0016	&	0.377	&	0.050	\\
	&	17	&	7981.63147	&	0.0012	&	0.291	&	0.044	\\
	&	17	&	7981.63147	&	0.0012	&	0.316	&	0.057	\\
 \hline
\end{longtable}
\end{center}

\begin{center}
\begin{longtable}{|l|l|ll|ll|}
\caption{Median values and 1-$\sigma$ limits of the posterior PDFs deduced for the timings and depths from their global analyses for K2 observations. Each row represents a transit, the first column gives the planet's name, the second the epoch of the transit, the third the mid-transit timing and the corresponding error resulting from the analysis and the last column the transit depth and corresponding error resulting from the analysis.} \label{global_k2} \\
\hline
\hline \multicolumn{1}{|c|}{\textbf{Planet}} & \multicolumn{1}{|c|}{\textbf{Epoch}} & \multicolumn{2}{|c|}{\textbf{Transit timing [$BJD_{TDB}-2450000$]}} & \multicolumn{2}{c|}{\textbf{Transit depth (\%)}} \\ \hline 
\endfirsthead

\multicolumn{3}{c}%
{{\bfseries \tablename\ \thetable{} -- continued from previous page}} \\

\endhead

\hline \multicolumn{3}{|r|}{{Continued on next page}} \\ \hline
\endfoot

\hline \hline
\endlastfoot

 \hline

b & 277 & 7741.02854 & 0.00088& 0.883 & 0.16 \\ 
 & 278 & 7742.54031 & 0.00100& 0.755 & 0.130 \\ 
 & 279 & 7744.05189 & 0.00060& 0.707 & 0.069 \\ 
 & 280 & 7745.56251 & 0.00069& 0.710 & 0.069 \\ 
 & 282 & 7748.58503 & 0.00073& 0.725 & 0.082 \\ 
 & 283 & 7750.09517 & 0.00130& 0.759 & 0.082 \\ 
 & 284 & 7751.60547 & 0.00093& 0.733 & 0.099 \\ 
 & 285 & 7753.11697 & 0.00093& 0.702 & 0.095 \\ 
 & 286 & 7754.62839 & 0.00068& 0.704 & 0.081 \\ 
 & 287 & 7756.13946 & 0.00095& 0.748 & 0.120 \\ 
 & 288 & 7757.64914 & 0.00096& 0.787 & 0.130 \\ 
 & 289 & 7759.16115 & 0.00095& 0.678 & 0.071 \\ 
 & 290 & 7760.67223 & 0.00092& 0.729 & 0.084 \\ 
 & 291 & 7762.18186 & 0.00067& 0.798 & 0.098 \\ 
 & 292 & 7763.69279 & 0.00130& 0.737 & 0.130 \\ 
 & 293 & 7765.20350 & 0.00056& 0.848 & 0.082 \\ 
 & 294 & 7766.71535 & 0.00058& 0.754 & 0.074 \\ 
 & 295 & 7768.22554 & 0.00086& 0.772 & 0.093 \\ 
 & 297 & 7771.24824 & 0.00150& 0.634 & 0.110 \\ 
 & 298 & 7772.75842 & 0.00120& 0.628 & 0.110 \\ 
 & 299 & 7774.26926 & 0.00093& 0.862 & 0.097 \\ 
 & 300 & 7775.78035 & 0.00099& 0.699 & 0.110 \\ 
 & 301 & 7777.28988 & 0.00067& 0.679 & 0.081 \\ 
 & 302 & 7778.80210 & 0.00086& 0.637 & 0.072 \\ 
 & 303 & 7780.31392 & 0.00089& 0.763 & 0.099 \\ 
 & 305 & 7783.33449 & 0.00099& 0.590 & 0.078 \\ 
 & 306 & 7784.84429 & 0.00200& 0.487 & 0.096 \\ 
 & 311 & 7792.40048 & 0.00060& 0.784 & 0.090 \\ 
 & 313 & 7795.42063 & 0.00095& 0.829 & 0.120 \\ 
 & 314 & 7796.93209 & 0.00087& 0.753 & 0.110 \\ 
 & 315 & 7798.44265 & 0.00078& 0.799 & 0.098 \\ 
 & 316 & 7799.95390 & 0.00090& 0.758 & 0.110 \\ 
 & 317 & 7801.46367 & 0.00093& 0.702 & 0.095 \\ 
 & 319 & 7804.48731 & 0.00062& 0.749 & 0.076 \\ 
 & 320 & 7805.99734 & 0.00120& 0.623 & 0.110 \\ 
 & 322 & 7809.01987 & 0.00050& 0.950 & 0.080 \\ 
 & 323 & 7810.52885 & 0.00070& 0.718 & 0.072 \\ 
 & 325 & 7813.55233 & 0.00087& 0.767 & 0.091 \\ 
 & 326 & 7815.06311 & 0.00069& 0.696 & 0.070 \\ 
 & 327 & 7816.57415 & 0.00014& 0.825 & 0.170 \\ 
 \hline
c & 189 & 7740.53434 & 0.00071& 0.572 & 0.057 \\
 & 190 &  7742.95387 & 0.00096& 0.711 & 0.085 \\
 & 191 &  7745.37552 & 0.00130& 0.602 & 0.079 \\
 & 192 &  7747.79788 & 0.00100& 0.772 & 0.099 \\
 & 193 &  7750.21885 & 0.00077& 0.685 & 0.058 \\
 & 194 &  7752.64222 & 0.00130& 0.620 & 0.069 \\
 & 196 &  7757.48369 & 0.00120& 0.713 & 0.110 \\
 & 197 &  7759.90363 & 0.00091& 0.542 & 0.087 \\
 & 198 &  7762.32938 & 0.00099& 0.662 & 0.091 \\
 & 199 &  7764.74912 & 0.00160& 0.736 & 0.096 \\
 & 200 &  7767.17049 & 0.00110& 0.741 & 0.076 \\
 & 201 &  7769.59284 & 0.00079& 0.549 & 0.075 \\
 & 202 &  7772.01581 & 0.01000& 0.823 & 0.072 \\
 & 203 &  7774.43569 & 0.00092& 0.681 & 0.068 \\
 & 204 &  7776.85852 & 0.00081& 0.715 & 0.060 \\
 & 205 &  7779.27989 & 0.00120& 0.674 & 0.090 \\
 & 206 &  7781.70123 & 0.00058& 0.768 & 0.060 \\
 & 207 &  7784.12346 & 0.00092& 0.795 & 0.089 \\
 & 210 &  7791.38893 & 0.00084& 0.589 & 0.081 \\
 & 211 &  7793.81172 & 0.00086& 0.657 & 0.081 \\
 & 212 &  7796.23247 & 0.00074& 0.746 & 0.078 \\
 & 214 &  7801.07714 & 0.00150& 0.734 & 0.120 \\
 & 215 &  7803.49838 & 0.00085& 0.624 & 0.078 \\
 & 216 &  7805.91962 & 0.00110& 0.606 & 0.110 \\
 & 217 &  7808.34096 & 0.00140& 0.744 & 0.082 \\
 & 219 &  7813.18461 & 0.00096& 0.641 & 0.065 \\
 & 220 &  7815.60652 & 0.00072& 0.825 & 0.064 \\
 \hline
d & 17 & 7738.99218 &0.00230& 0.258 & 0.065 \\
 & 18 & 7743.03815 &0.00087& 0.562 & 0.091 \\
 & 20 & 7751.14085 &0.00230& 0.434 & 0.079 \\
 & 21 & 7755.18922 &0.00130& 0.428 & 0.072 \\
 & 22 & 7759.24736 &0.00210& 0.441 & 0.070 \\
 & 23 & 7763.28937 &0.00140& 0.408 & 0.065 \\
 & 24 & 7767.33969 &0.00260& 0.283 & 0.070 \\
 & 26 & 7775.44044 &0.00160& 0.454 & 0.082 \\
 & 30 & 7791.64168 &0.00088& 0.549 & 0.062 \\
 & 36 & 7815.94088 &0.00260& 0.289 & 0.070 \\
 \hline
e & 13 & 7739.67188& 0.00610&0.478 & 0.089 \\
 & 14 & 7745.77245& 0.00430&0.473 & 0.072 \\
 & 16 & 7757.96794& 0.00340&0.572 & 0.120 \\
 & 17 & 7764.06998& 0.00120&0.477 & 0.077 \\
 & 18 & 7770.17137& 0.00270&0.413 & 0.071 \\
 & 19 & 7776.26467& 0.00190&0.365 & 0.063 \\
 & 20 & 7782.36298& 0.00170&0.414 & 0.059 \\
 & 22 & 7794.56266& 0.00210&0.587 & 0.092 \\
\hline
f & 8 & 7745.03110& 0.00230& 0.567 & 0.090 \\
 & 9  & 7754.23467& 0.00160&0.603 & 0.069 \\
 & 10 & 7763.44538& 0.00200&0.636 & 0.100 \\
 & 11 & 7772.64872& 0.00220&0.456 & 0.070 \\
 & 14 & 7800.27402& 0.00230&0.494 & 0.088 \\
 & 15 & 7809.47707& 0.00170&0.484 & 0.064 \\
 \hline
g & 8 & 7764.19196& 0.00160& 0.567 & 0.068 \\
 & 11 & 7801.25070& 0.00120&0.707 & 0.087 \\
 & 12 & 7813.60635& 0.00140&0.728 & 0.100 \\
 \hline
h & 5 & 7756.38806& 0.00300&0.346 &	0.058 \\
\label{appendix:global_K2}
\end{longtable}
\end{center}

\begin{center}
\begin{longtable}{|l|l|ll|ll|}
\caption{Median values and 1-$\sigma$ limits of the posterior PDFs deduced for the timings and depths from their global analyses for Liverpool telescope observations. Each row represents a transit, the first column gives the planet's name, the second the epoch of the transit, the third the mid-transit timing and the corresponding error resulting from the analysis and the last column the transit depth and corresponding error resulting from the analysis.}  \\
\hline
\hline \multicolumn{1}{|c|}{\textbf{Planet}} & \multicolumn{1}{|c|}{\textbf{Epoch}} & \multicolumn{2}{c|}{\textbf{Transit timing [$BJD_{TDB}-245000$]}} & \multicolumn{2}{c|}{\textbf{Transit depth (\%)}} \\ \hline 
\endfirsthead

\multicolumn{3}{c}%
{{\bfseries \tablename\ \thetable{} -- continued from previous page}} \\
\endhead

\hline \multicolumn{3}{|r|}{{Continued on next page}} \\ \hline
\endfoot

\hline \hline
\endlastfoot
\hline
b & 386 & 7905.71519  & 0.00088 & 0.834 & 0.120  \\
& 421 & 7958.59605  & 0.00036 & 0.687 & 0.061  \\
& 425 & 7964.63885  & 0.00044 & 0.838 & 0.053  \\
& 429 & 7970.68541  & 0.00041 & 0.707 & 0.062  \\
\hline
  c & 270 & 7936.69651  & 0.00035 & 0.723 & 0.047  \\
    & 298 & 8004.50488  & 0.00053 & 0.853 & 0.054  \\
    & 303 & 8016.61367  & 0.00068 & 0.605 & 0.084  \\
    & 319 & 8055.36297  & 0.00047 & 0.764 & 0.066   \\
    & 284 & 7970.60044  & 0.00088 & 0.641 & 0.070   \\
\hline
d & 86 & 8018.43071  & 0.00096 & 0.353 & 0.027  \\
\hline
 e & 53 & 7983.62906  & 0.00130 &  0.476 & 0.069  \\
 & 56 & 8032.43405  & 0.00190 & 0.478 & 0.100  \\
\hline
h & 17 & 7981.63343   & 0.00110 & 0.257  & 0.035  \\
 \label{global_liverpool}
\end{longtable}
\end{center}

\newpage
\label{AppendixD}
\section{Combined transit depth values from observation versus Z18's predictions}

\begin{table*}[h!]
\centering                          
\begin{tabular}{c c c c c}        
\hline\hline                 
Planets & \textbf{Effective bandpass} ($\mu$m) & Z18 (\%) & Observations (\%)  \\    
\hline  
 b+c  & 4.5  & 1.44 $\pm$ 0.03 & 1.424 $\pm$ 0.008 \\      
 &	1.6	& 1.54 $\pm$ 0.03 &		1.539	$\pm$	0.028	\\
 &	1.55	& 1.52 $\pm$ 0.03&		1.536	$\pm$	0.033	\\
 &	1.5	& 1.49 $\pm$ 0.03&		1.542	$\pm$	0.033	\\
 &	1.45	& 1.45 $\pm$ 0.03&		1.534	$\pm$	0.040	\\
 &	1.4	& 1.42$\pm$ 0.03 &		1.494	$\pm$	0.037	\\
 &	1.35	&	1.46 $\pm$ 0.03&	1.484	$\pm$	0.034	\\
 &	1.3	&	1.51 $\pm$ 0.03&	1.534	$\pm$	0.035	\\
 &	1.25	&	1.54$\pm$ 0.03 &	1.592	$\pm$	0.033	\\
 &	1.2	& 1.53$\pm$ 0.03 &		1.531	$\pm$	0.028	\\
 &	1.15	& 1.53$\pm$ 0.03 &		1.487	$\pm$	0.039	\\
  & 0.8-1.1 &  1.33 $\pm$ 0.03 & 1.470  $\pm$ 0.032 \\
  & 0.73-1.1&  1.27 $\pm$ 0.03 & 1.490 $\pm$ 0.027 \\
  & 0.55-0.9 & 0.94 $\pm$ 0.03 & 1.400 $\pm$ 0.020 \\
  \hline
 b+c+d+e+f+g  & 4.5  & 3.55 $\pm$ 0.06 & 3.646 $\pm$ 0.009 \\   
 &	1.63	&  3.91 $\pm$ 0.06  &	3.885	$\pm$	0.027	\\
 &	1.58	&  3.72 $\pm$ 0.06 &	3.873	$\pm$	0.032	\\
 &	1.53	&  3.75 $\pm$ 0.06 &	3.793	$\pm$	0.032	\\
 &	1.48	&  3.78 $\pm$ 0.06 &	3.824	$\pm$	0.032	\\
 &	1.43	&  3.47 $\pm$ 0.06 &	3.750	$\pm$	0.035	\\
 &	1.38	&  3.79 $\pm$ 0.06 &	3.759	$\pm$	0.033	\\
 &	1.33	&  3.86 $\pm$ 0.06 &	3.858	$\pm$	0.038	\\
 &	1.28	&  3.89 $\pm$ 0.06 &	3.895	$\pm$	0.03	\\
 &	1.23	&  3.89 $\pm$ 0.06 &	3.834	$\pm$	0.029	\\
 &	1.18	&  3.88 $\pm$ 0.06 &	3.771	$\pm$	0.033	\\
  & 0.8-1.1 & / & / \\
  & 0.73-1.1 & 3.34  $\pm$ 0.06 & 4.370    $\pm$  0.049   \\
  & 0.55-0.9  &  2.62 $\pm$ 0.06 & 3.474 $\pm$ 0.038 \\
  \hline
 d+e+f+g  & 4.5  & 2.19 $\pm$ 0.05 & 2.222 $\pm$ 0.010 \\      
  &	1.63	& 2.37 $\pm$ 0.05  &	2.345	$\pm$	0.023	\\
 &	1.58	& 2.27 $\pm$ 0.05 &	 2.337	$\pm$	0.027	\\
 &	1.53	& 2.28 $\pm$ 0.05 &	 2.251	$\pm$	0.027	\\
 &	1.48	& 2.29 $\pm$ 0.05 &	 2.291	$\pm$	0.025	\\
 &	1.43	& 2.13 $\pm$ 0.05 &	2.257	$\pm$	0.029	\\
 &	1.38	& 2.30 $\pm$ 0.05 &	2.276	$\pm$	0.028	\\
 &	1.33	& 2.34 $\pm$ 0.05 &	2.324	$\pm$	0.033	\\
 &	1.28	& 2.35 $\pm$ 0.05 &	2.303	$\pm$	0.025	\\
 &	1.23	& 2.35 $\pm$ 0.05 &	2.303	$\pm$	0.026	\\
 &	1.18	& 2.35$\pm$ 0.05 &	2.284	$\pm$	0.027	\\
  & 0.8-1.1 & / &  / \\     
  & 0.73-1.1 & 2.05 $\pm$ 0.05 & 2.233 $\pm$ 0.037 \\
  & 0.55-0.9 & 1.66 $\pm$ 0.05 & 2.074 $\pm$ 0.044 \\
\hline                                   
\end{tabular}

\caption{Combined transit depth values (in percent) for b+c, b+c+d+e+f+g, and d+e+f+g, as predicted from the best-fit stellar contamination model of Z18, and as measured from K2, SPECULOOS, HST/WFC3, and {\it Spitzer} observations in their effective bandpass relatively to an M8 star spectrum}.
\label{depth_value_combined}
\end{table*}

\begin{figure}[h!]
\centering
\includegraphics[angle=0,width=12cm]{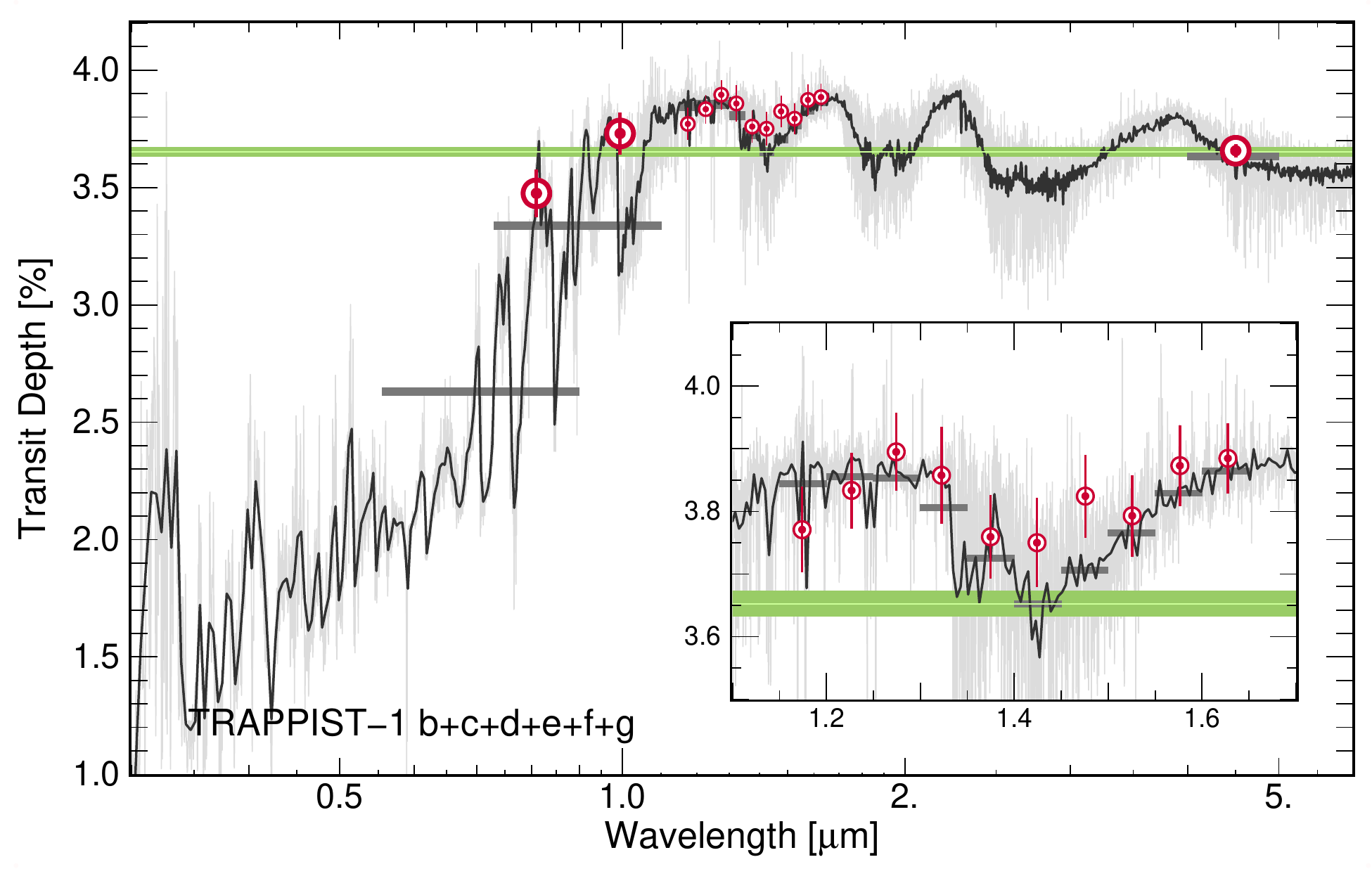}
\caption{Comparison of the stellar contamination spectrum inferred by Z18 for TRAPPIST-1\,b+c+d+e+f+g transits [\cite{Zhang2018}] at two different resolutions (continuous black line and gray line) with the K2 and SSO measurements presented in this work, and the {\it Spitzer} and HST/WFC3 presented in \cite{Delrez2018} and \cite{deWit2016}, respectively (red points). The green line represents the weighted mean of all measurements except HST for the reasons outlined earlier in Section. \ref{Transmission_spectra}. Finally, the gray horizontal bars are the band-integrated value for the Z18 model where the integrals are weighted uniformly in wavelength.
}
\color{black}
\label{Zhang_vs_obs_bcdefg}
\end{figure}

\newpage
\section{Limb-darkening}
\label{AppendixE}

\begin{table}[h!]
\centering                          
\begin{tabular}{c c c c}        
\hline\hline                 
 
Telescope	& Planet  &   $dF_{LD}$ (\%) &  $dF_{Analyses}$ (\%) \\
\hline
K2	& 1b  &    0.751 $\pm$ 0.027 & 0.716 $\pm$ 0.021 \\
& 1c  &    0.712 $\pm$ 0.009 & 0.684 $\pm$ 0.019 \\
& 1d  &    0.386 $\pm$ 0.009 & 0.412 $\pm$ 0.028 \\
& 1e  &    0.460 $\pm$ 0.009 & 0.449 $\pm$ 0.034 \\
& 1f  &    0.617 $\pm$ 0.067 & 0.541 $\pm$ 0.034 \\
& 1g  &    0.741 $\pm$ 0.026 & 0.668 $\pm$ 0.070 \\
& 1h  &    0.291 $\pm$ 0.029 & 0.347 $\pm$ 0.058 \\
\hline
\end{tabular}
\caption{Comparison of $dF_{Analyses}$, the transit depth values obtained from a global analysis of all the K2 transits for each planet, with $dF_{LD}$, the transit depth values obtained from a global analysis of the period-folded TTV-corrected K2 transit photometry with free
%
%
limb-darkening coefficients  for all planets.}  
\label{limb_darkening}
\end{table}

\newpage
\bibliographystyle{achemso}
\bibliography{biblio}
\end{document}